\documentclass[%
aip,
sd,%
amsmath,amssymb,
author-numerical,%
]{revtex4-1}

\usepackage{graphicx}% Include figure files
\usepackage{dcolumn}% Align table columns on decimal point
\usepackage{bm}% bold math
\usepackage[dvipsnames]{xcolor}
\usepackage[normalem]{ulem}
\usepackage[title]{appendix}
\usepackage{comment}

\newcommand{\FS}[1]{{\color{black}{#1}}}
\graphicspath{{./figures/}}
\newtheorem{definition}{Definition}
\begin{document}
	
	%\preprint{AIP/123-QED}
	
	\title{Stability Analysis of Reservoir Computers Dynamics via Lyapunov Functions}% Force line breaks with \\
	%\thanks{Footnote to title of article.}

	\author{Afroza Shirin}
	\email{ashirin@unm.edu}
	\affiliation{ 
		Mechanical Engineering Department, University of New Mexico, Albuquerque, NM 87131}
	
	\author{Isaac S. Klickstein}
	\email{iklick@unm.edu}
	\affiliation{ 
		Mechanical Engineering Department, University of New Mexico, Albuquerque, NM 87131}

	\author{Francesco Sorrentino}
	\email{fsorrent@unm.edu}
	\affiliation{
		Mechanical Engineering Department, University of New Mexico, Albuquerque, NM 87131}
	
	\date{\today}% It is always \today, today,
	
	\begin{abstract}
		A Lyapunov design method is used to analyze  the nonlinear stability of a generic reservoir computer for both the cases of continuous-time and discrete-time dynamics. Using this method, for a given nonlinear reservoir computer,  a radial region of stability  around a fixed point is analytically determined. \FS{We see that the training error of the reservoir computer is lower in the region where the analysis predicts global stability but is also affected by the particular choice of the individual dynamics for the reservoir systems. For the case that the dynamics is polynomial, it appears to be important for the polynomial to have nonzero coefficients corresponding to at least one odd power (e.g., linear term) and one even power (e.g., quadratic term).}	%
	\end{abstract}
	
	%Use showkeys class option if keyword
	\maketitle
	
	\begin{quotation}
		\FS{
		While nonlinearity appears to be a fundamental component of reservoir computers, not much research has been performed to analyze stability of the nonlinear dynamics of these systems.  
		In this paper, we use a Lyapunov design method to estimate the basin of attraction of a fixed point for the dynamics of a generic reservoir computer. Our nonlinear stability analysis unveils a trade-off between the need for global stability, which is achievable by linear dynamics alone, and the need for higher-order terms of the dynamics, which could in turn compromise stability. %Our analysis empowers us to design an optimization problem to find the optimal set of parameters that optimize the performance of the reservoir computer.
	  }
	\end{quotation}
	
\section{\label{sec:level1}Introduction}
\indent
A reservoir computer (RC) is a complex nonlinear dynamical system that is used for processing and analyzing empirical data, see e.g. \cite{jaeger2001echo, schrauwen2007overview, natschlager2002liquid, maass2002real, martinenghi2012photonic,brunner2013parallel,nakajima2015information, hermans2015photonic,vinckier2015high,duport2016fully,larger2017high}, 
modeling of complex dynamical systems \cite{suykens2012artificial}, 
speech recognition \cite{crutchfield2010introduction},  learning of context free and context sensitive languages \cite{rodriguez2001simple, gers2001lstm},
the reconstruction and prediction of chaotic attractors \cite{lu2018attractor, zimmermann2018observing,antonik2018using, jaeger2004harnessing,pathak2017using,pathak2018model},
image recognition \cite{jalalvand2018application}, and
control of robotic systems \cite{graves2004biologically, robinson1994application,lukovsevivcius2012reservoir}. 

A typical RC consists of a set of nodes coupled together to form a network. Each node of the RC evolves in time in response to an input signal that is externally provided to the reservoir. An output signal is then generated from the time evolutions of the RC nodes.
In a RC, the output connections (those that connect the RC nodes to the output) are trained to produce a best fit between the output signal and a training signal related to the original input signal. On the other hand, the connections between the nodes of the reservoir are constant parameters of the system. As a result, RCs are easier to analyze than other machine learning tools for which all the connections are typically  trained.

The functions of RCs mainly depend on two factors; (i)  nonlinearity of the nodal dynamics which is needed to process the information in the input signal and (ii) linear memory to boost the excitability of the RC dynamics \cite{dambre2012information}. Though earlier works have shown that maximizing linear memory is important in  information processing \cite{jaeger2002short,ganguli2008memory,buonomano2009state,white2004short}, more recent works have shown that the performance of a reservoir computer is related to consideration of both factors (i) and (ii) \cite{verstraeten2007experimental,inubushi2017reservoir,marzen2017difference}. In addition, the performance of a reservoir computer is also affected by a number of other factors, including  the  reservoir adjacency matrix, i.e., the strengths of the connections between the RC nodes, and the  dynamic range of the input signals \cite{verstraeten2009quantification}. %All of the above factors affect the performance of reservoir computers, and the local gain, {which is defined as the ratio between the magnitude of the input signal and the magnitude of the output signal at steady state.} 

From linear systems theory, a dynamical system is reliable and safe when it is stabilizable around  some operating point \cite{sontag2013mathematical}. Previous research has used linear stability to assess the stability of RCs around this operating point \cite{marcus1989stability,tanaka1995stability,belair1996frustration}. However, as the RC dynamics requires nonlinearity for its proper operation, a linear stability analysis of the RC dynamics around a specific point is not sufficient. This motivates us to develop a nonlinear stability analysis, based on Lyapunov functions, which we use to characterize the basin of attraction of the desired operating point. As assessing stability of the nonlinear system when forced by an external stimulus is contingent on the particular stimulus provided, we characterize nonlinear stability of the unforced RC dynamics. If the desired operating point is found to be globally asymptotically stable, then stability is independent of the particular stimulus, as long as it is bounded.

%The basin of attraction  is typically sensitive to changes in the parameters of the reservoir computer. 

We compute a constant $c$-radius region around the operating point such that the dynamics of the system remains bounded inside this region. This can be done by choosing the parameters of the system and the type of nonlinearity, with the goal of possibly enhancing the performance of the reservoir computer. We consider different types of nonlinear dynamics at the network nodes, e.g., polynomial, in either continuous time or discrete time.  This differs from the common approach in the literature, where the nodal dynamics is chosen to have a {squashing} type nonlinearity  (in most cases a sigmoid function, e.g., $\tanh()$)  so that the states of the nonlinear system always remain bounded inside some region  \cite{liu2001stability,saul2000attractor,yang2014exponential}. {(A squashing function is defined as a function that is monotonic and bounded within a small range. For example, the function $\tanh()$ squashes the argument to the interval $[-1,1]$.)}

Few theoretical works have investigated  how the underlying stability of a nonlinear RC  affects its performance. Reference\ \cite{ganguli2008memory} showed that the total memory capacity, a measure of performance, is related to the size of the network. This analysis did not consider how the underlying stability of the system is related to the total memory capacity.
In Ref.\ \cite{verstraeten2010memory}, an optimal parameter setting for the reservoir was studied, but no direct relation between  the dynamic properties in terms of nonlinear stability and the reservoir performance was provided.
Previous work \cite{verstraeten2009quantification} found that the performance of the RC was improved when the condition number of the Jacobian of the reservoir dynamics was small.
The references listed herein, and others, motivated us to perform a rigorous dynamical investigation of the nonlinear stability of RCs.
%Previous work \cite{verstraeten2009quantification} studied the role of the RC dynamics in affecting their  performance and found that improved performance is achieved when the ratio between the largest singular value and the smallest singular value of the Jacobian of the stable reservoir dynamics is small. This and other papers motivated us to perform a rigorous dynamical investigation of the nonlinear stability of RCs.

In this paper, for a given nonlinear RC, we determine the $c (\bm{\theta})$-region of stability, %for which it performs better than any other settings of the parameters. Note that
where $c$ is the radius around a fixed point and $\bm{\theta}$ is the set of  parameters of the reservoir.
We use Lyapunov design methods \cite{haddad2011nonlinear} to find the  $c(\bm{\theta})$-region  where a nonlinear reservoir computer is stable, safe, reliable and its  performance is {predictable}. Lyapunov design methods are used widely in controls engineering to design controllers that achieve qualitative objectives, such as stabilizing a system or maintaining a system’s state in a desired operating range \cite{sontag2013mathematical}. To the best of our knowledge,  the use of a Lyapunov-based design approach with respect to the performance of a reservoir computer is novel. In Ref.\ \cite{perkins2002lyapunov}, a Lyapunov function has been used to design the controller of a reinforcement learning system, but the paper does not show how stability {of the RC}  affects its performance. In this article, first we assume that the input signal is normalized and scaled properly, the eigenvalues of the adjacency matrix satisfy certain constraints, and second we design the $c(\bm{\theta})$-region of stability by using a Lyapunov design method. Our nonlinear stability analysis provides insight into the effects of the RC parameters on its performance.

{In Sec.\ II we lay out the general theory to assess the basin of attraction within which the RC dynamics is stable for both  continuous-time dynamics and discrete-time dynamics. In Sec.\ III we %present the results of our numerical calculations that we performed to 
investigate the relation between our stability predictions and the RC performance  in terms of the computed training error. Finally, conclusions are presented in Sec.\ IV.}

%% Include Methods Section
\section{\label{sec:level3}Methods}
%\input{sections/methods}
%%  Methods
\subsection{Reservoir Computer with Continuous Time Dynamics}
We consider the dynamics of a reservoir computer as continuous-time \cite{carroll2019network},
\begin{equation}\label{eq:cont:dyn1}
\dot{r}_i(t) =  f(r_i(t),\bm{\theta}) + \sum_{j = 1}^{ M} A_{ij} r_j(t) + w_i s(t), \quad i = 1,2,\cdots, M
\end{equation}
and the unforced (without input) reservoir dynamics, 
\begin{equation}\label{eq:cont:dyn2}
\dot{r}_i(t) =  f(r_i(t),\bm{\theta}) + \sum_{j = 1}^{M} A_{ij} r_j(t), \quad i = 1,2,\cdots, M
\end{equation}
where $r_i(t) \in \mathbb{R}, i = 1,2,\cdots,M$ denotes the state of node $i$ at time $t \in \mathbb{R}$, $f(r_i,\bm{\theta})$ is the nonlinear \emph{nodal dynamics}  at node $i$, the adjacency matrix $A=\{A_{ij}\}$ indicates the coupling from node $j$ to node $i$, $s(t) \in \mathbb{R}$ is an input signal and $\textbf{w} = \left[w_1 , w_2,\cdots,w_M \right]$ is a vector describing the coupling of input signal $s(t)$ with node $i$. We assume that $r_i^*(t) = 0$ is a (linearly) stable fixed point of the system in Eq.\ \eqref{eq:cont:dyn2} and the input signal $s(t)$ in Eq.\ \eqref{eq:cont:dyn1} is normalized to have zero mean  and standard deviation equal to one.

\subsubsection{Lyapunov Function and $c(\bm{\theta})$-region design}
%We consider the unforced dynamics in Eqs. \eqref{eq:cont:dyn2}. 
We define a Lyapunov function $V : \mathcal{D}(c) \subseteq \mathbb{R}^M \rightarrow \mathbb{R}$ for the unforced dynamical equation in Eq. \eqref{eq:cont:dyn2},
\begin{equation}\label{eq:cont:lyap}
V(\textbf{r})= \frac{1}{2} \textbf{r}^T\textbf{r},
\end{equation}
where $\mathcal{D}(c) = \left\{ \textbf{r} = (r_1,r_2,\cdots,r_M ) \in \mathbb{R}^M :  \|\textbf{r}\| \le c \right\}$ is the phase space region that is included in a hypersphere of radius $c$, centered at the origin. Here $V(\textbf{r})>0$ for $\textbf{r} \ne 0$ and $V(\textbf{0})= 0$. Then for all $\textbf{r} \in \mathcal{D}(c)\backslash \textbf{0} $, 
\begin{subequations}
	\begin{align}
	\frac{\partial V}{\partial {\textbf{r}}} \textbf{f}(\textbf{r},\bm{\theta}) =  & \textbf{r}^T \left(  \textbf{f}(\textbf{r}(t),\bm{\theta}) + A\textbf{r} \right) \\ 
	= & \sum_i r_if(r_i, \bm{\theta})  +\textbf{r}^T A\textbf{r} \\
	=	& \sum_i r_if(r_i, \bm{\theta})  + \frac{1}{2} \textbf{r}^T (A+A^T)\textbf{r} \\
	=	& \sum_i r_if(r_i, \bm{\theta})  +  \textbf{r}^T A_{s} \textbf{r}\\
	\le 	& \sum_i r_if(r_i, \bm{\theta})  +  \alpha_{\max} \|\textbf{r}\|^2
	\label{eq:cont:lyap-in}
	\end{align}
\end{subequations}
where $A_s $ is the symmetric part of the matrix $A$, $\alpha_{\max}$ is the largest real eigenvalue of the matrix $A_s$, and $\bm{\theta}$ is the set of parameters that completely characterize the nonlinear function $f$.
Now we introduce a quadratic upper bound to  the term $r_if(r_i,\bm{\theta})$. Let us consider that the term $K(c, \bm{\theta}) r_i^2$ is such a quadratic upper bound, that is $r_if(r_i,\bm{\theta}) \le r_i^2 K(c,\bm{\theta})$, where  $K(c, \bm{\theta})$ is a scalar function of $c$ and $\bm{\theta}$.
The inequality in Eq.\ \eqref{eq:cont:lyap-in} can now be written as,
\begin{subequations}
	\begin{align}
	& \sum_i r_if(r_i, \bm{\theta})  +  \alpha_{\max} \|\textbf{r}\|^2\\
	\le & \sum_i  K(c,\bm{\theta})r_i^2  +\alpha_{\max} \|\textbf{r}\|^2\\
	= & \left( K(c,\bm{\theta}) + \alpha_{\max} \right) \|\textbf{r}\|^2
	\end{align}
\end{subequations}
According to the second Lyapunov stability theorem \cite{haddad2011nonlinear}, the system in Eq.\ \eqref{eq:cont:dyn2} is stable within  $\mathcal{D}$ if the following inequality holds,
\begin{subequations} \label{eq:ineq}
	\begin{align}
	& \left( K(c,\bm{\theta}) + \alpha_{\max} \right) \|\textbf{r}\|^2 \le 0 &&\\
	\text{or equivalently, } \quad  & K(c,\bm{\theta}) \le -  \alpha_{\max} &&
	\end{align}
\end{subequations}
Note that for each $c$, the term on the left hand side of Eq.\ (\ref{eq:ineq}b) depends on the dynamics and parameters of the individual nodes and the term on the right hand side of Eq.\ (\ref{eq:ineq}b) depends on the network topology. Thus  Eq.\ (\ref{eq:ineq}b) effectively decouples the stability problem into two terms that can be adjusted independently of each other: the nodal dynamics and the network topology.

We now provide a definition of $c(\bm{\theta})$-region stability of a reservoir computer.
\begin{definition}
	A nonlinear reservoir computer is $c(\bm{\theta})$-region stable if $K(c,\bm{\theta}) \le -\alpha_{\max} $. 
\end{definition}
Note that this is a sufficient (but not necessary) condition for a reservoir to be stable inside the region $\mathcal{D}(c)$. Also, if $c \rightarrow \infty$, the system is globally asymptotically stable. 

We note that for any constant  $K\le K(c,\bm{\theta})$, the system is $c(\bm{\theta})$-region stable. We will thus attempt to find an upper bound $K(c,\bm{\theta})r_i^2$ to $r_i f(r_i ,\bm{\theta})$ that is as tight as possible. To find the minimal $K\le K(c,\bm{\theta})$, we define an optimization problem,
\begin{subequations}\label{eq:cont:opt}
	\begin{align}
	\min_K  & && K(c,\bm{\theta})\\
	\text{ s.t. } &  && r_if(r_i) -	Kr_i^2 \le 0, \quad \forall r_i \in [-c,c]
	\end{align}
\end{subequations}
As $r_i$ lies within the closed interval $[-c,c]$, the constraint in Eq.\ \eqref{eq:cont:opt} must be satisfied along a continuum. However, we know that the constraint in Eq. (\ref{eq:cont:opt}b) achieves equality for some $r_i^* \in [-c,c]$  at the optimal solution $K^*(c,\bm{\theta})$,
\begin{equation}
\begin{aligned}
K^*(c,\bm{\theta}){r^*_i}^2 - r_i^*f(r^*_i,\bm{\theta}) & = 0 
\end{aligned}
\end{equation} 
or equivalently,
\begin{equation}
\begin{aligned}
K^*(c,\bm{\theta}) & = \frac{f(r^*_i,c)}{r^*_i} \ge  \frac{f(r_i,c)}{r_i}, \quad \forall r_i \in [-c,c]
\end{aligned}
\end{equation} 
The optimal coefficient $K^*(c,\bm{\theta})$ is chosen as the term of maximum value in the set,
\begin{equation}\label{eq:cont:kstar}
K^*(c,\bm{\theta}) = \max 		
\begin{cases}
\frac{f(c,\bm{\theta})}{c},\\
\frac{f(-c,\bm{\theta})}{-c},\\
f'(0,\bm{\theta}),\\
\frac{f(r^*_i,\bm{\theta})}{r_i^*}, \text{ where } r^*_i \text{ is the solution of }\\
\quad \quad \quad r^*_if'(r^*_i,\bm{\theta})-f(r^*_i,\bm{\theta}) = 0, \quad  r^*_i \in[-c,c], r_i^* \ne 0 
\end{cases}
\end{equation} 
where the four cases in Eq. \eqref{eq:cont:kstar} are the possible maxima of the ratio $f(r_i,c)/r_i$ over a closed interval.
The stability condition for the reservoir computer is
\begin{equation}\label{eq:cont:ineq}
K^*(c,\bm{\theta}) \le - \alpha_{\max}		
\end{equation}
From the inequality in Eq.\ \eqref{eq:cont:ineq} and the definition of $K^*(c,\boldsymbol{\theta})$ in Eq. \eqref{eq:cont:kstar}, we can find $c_{\max}$ for which $K^*(c_{\max}, \bm{\theta}) = - \alpha_{\max}$ and the  $c(\bm{\theta})$-region is determined by $\mathcal{D}(c_{\max}) = \{ \textbf{r} = (r_1,r_2,\cdots,r_M) \in \mathbb{R}^M : \|\textbf{r}\| \le c_{\max} \}$. We call $c_{\max}$ the radius of the region $\mathcal{D}(c_{\max})$. 
If $\lim\limits_{c \rightarrow \infty} K^*(c,\boldsymbol{\theta}) < -\alpha_{\max}$ then we say the system is globally stable (less formally we say $c_{\max} = \infty$), while if $K^*(0,\boldsymbol{\theta}) > -\alpha_{\max}$ then the system is unstable.
%If $c_{max} \rightarrow \infty$, the system is globally stable, and if $c_{max} \rightarrow 0$, the system is unstable.  We note that the region $\mathcal{D}(c_{\max})$ is contained inside the basin of attraction.
In the next subsection, we will find the  $c(\bm{\theta})$-region for the case that the nonlinear nodal dynamics is described by a polynomial.

\subsubsection{Polynomial Type Nonlinearity}

We now consider an example for which the reservoir computer consists of $M$ homogeneous nodes and the nodal dynamics of each node $i, i = 1,2,\cdots,M$ is defined by the following third-order polynomial function \cite{carroll2019network,carroll2019mutual}, 
\begin{equation}
f(r_i,\bm{\theta}) = p_1r_i+p_2r_i^2 +p_3 r_i^3,
\end{equation}
Polynomial functions are a very general way to express nonlinearity.

The dynamical equation that governs the evolution of each node $i$ is,
\begin{equation}\label{eq:cont:poly}
\dot{r}_i(t)  =  p_1r_i+p_2r_i^2 +p_3 r_i^3 + \sum_{j = 1}^M A_{ij}r_j + w_i s(t).
\end{equation}
Here, $p_1$, $p_2$ and $p_3$ are the coefficients of the polynomial. 
In this case, the set of parameters $\bm{\theta} = \{p_1,p_2,p_3\}$. The origin $r_i=0$ is a fixed point for the dynamics and is linearly stable if the largest real part of the eigenvalues of the matrix $(A -p_1 I)$ is negative.
Therefore, in what follows, we fix $p_1$ so as to ensure that the matrix $(A -p_1 I)$ is Hurwitz and then we characterize the basin of attraction as a function of the remaining parameters $p_2$ and $p_3$.

According to Eq.\ \eqref{eq:cont:kstar}, the  scalar function $K^*(c,\bm{\theta})$ can be obtained as,
\begin{equation}
\begin{aligned}
K^*(c,\bm{\theta}) &  = \max 		
\begin{cases}
\frac{f(c,\bm{\theta})}{c} = p_1+p_2c +p_3c^2,   \\
\frac{f(-c,\bm{\theta})}{-c} = p_1-p_2c +p_3c^2, \\
f'(0,\bm{\theta}) =  p_1, &\\
\frac{f(r^*_i,\bm{\theta})}{r^*_i} = (p_1 - \frac{p_2^2}{4p_3}), \text{ where } r^*_i =   \frac{-p_2}{2p_3} \in [-c,c], ,  r^*_i \ne 0  \\
\end{cases} &&&&&&& \\
\end{aligned}
\end{equation} 
We note that the condition  $K^*(c,\bm{\theta}) = -\alpha_{\max}(A_s)$ determines the radius $c_{\max}$ of the sphere $\mathcal{D}$ for which the reservoir computer is $c$-region stable.  Global stability is achieved when $K^*(c,\bm{\theta})$ remains upper bounded by $-\alpha_{\max}(A_s)$ as $c \rightarrow \infty$. 

\begin{figure}
	\centering
	\includegraphics[width=1\textwidth]{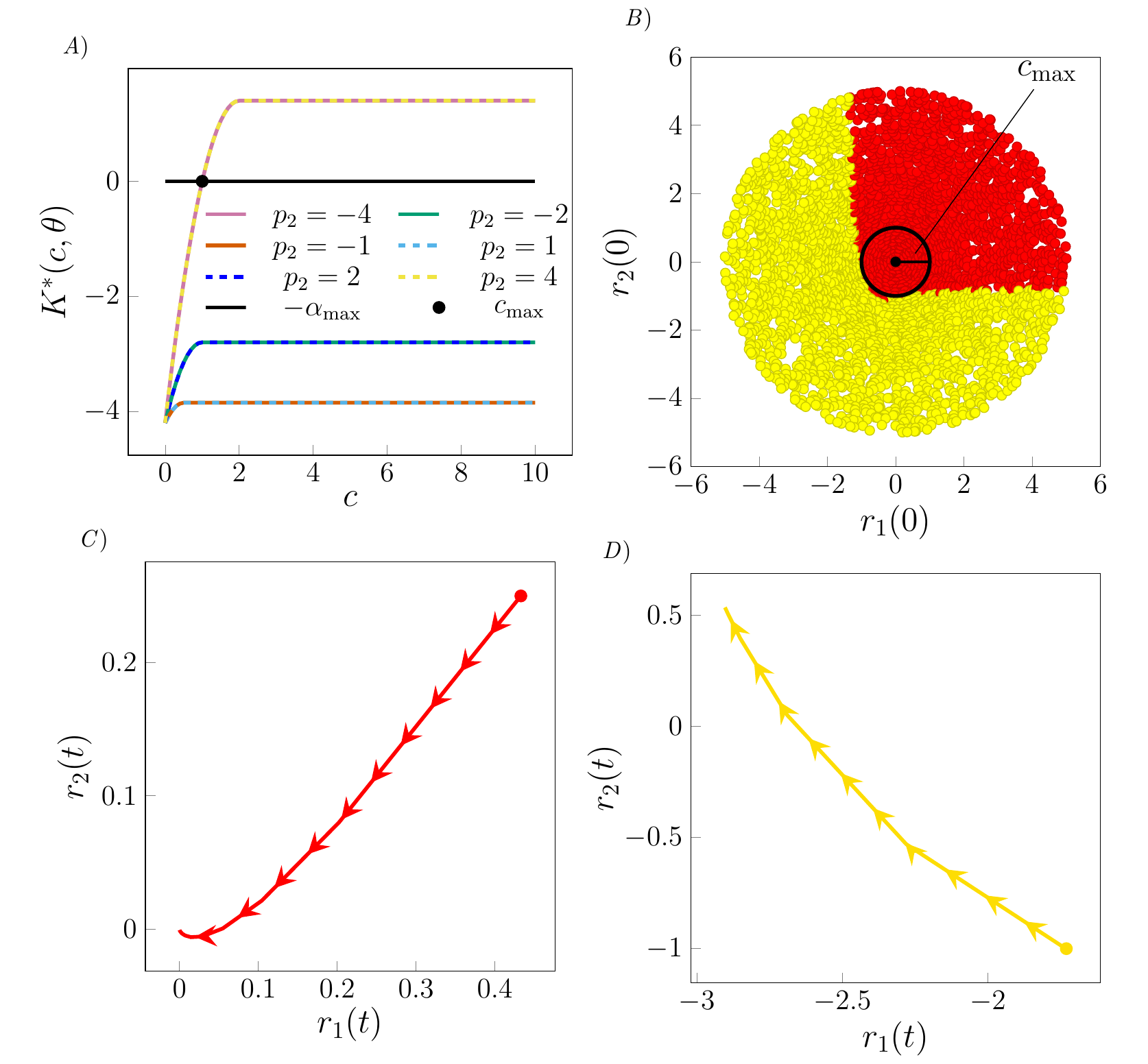}
	\caption{{\bf{Illustration of the $c$-region of stability for a simple reservoir computer of 2 nodes with polynomial type dynamics.} }
		(\emph{A}) The scalar function $K^*(c,\bm{\theta})$ versus $c$ for different values of $p_2$ where $p_1 = -3$ and $p_3 = -1$ are fixed. The black solid line is the constant-ordinate line at $-\alpha_{\max}$. The black dot represents the point at which $K^*(c_{\max},\bm{\theta}) = -\alpha_{\max}$ for $p_2 = \pm 4$. 
		(\emph{B}) The red region indicates the  initial conditions from which the unforced system evolution  approaches to the origin for large time. The yellow region indicates the initial conditions from which the system time evolution does not converge to the origin. The black circle represents $||\textbf{r}|| = c_{max}$ for $p_2 = \pm 4$ and tightly fits inside the numerically computed red region. 
		(\emph{C})  The trajectory  $r_2(t)$ versus $r_1(t)$ when the system is evolved from a typical initial condition from the red region. 
		(\emph{D})   The trajectory  $r_2(t)$ versus $r_1(t)$ when the system is evolved from a typical initial condition from the yellow region. In C and D arrows point in the direction of increasing time.}
	\label{fig:cont:example}
\end{figure}

Here, we provide an example to explain how to find the $c(\bm{\theta})$-region for a simple reservoir computer with $M=2$ nodes. We set $p_1 = -3$, $p_3 = -1$, $A = \begin{bmatrix} 0  &1 \\ -1 & 0 \end{bmatrix}$ and let the parameter $p_2$ vary. In Fig.\ \ref{fig:cont:example}(\emph{A}), we plot $K^*(c,\bm{\theta})$ versus $c$ for different values of $p_2$. The solid black line is the constant-ordinate line at $-\alpha_{\max}=0$. For this example, we observe that $K^*(c,\bm{\theta})$  is symmetric about the parameter $p_2 = 0$. For $p_2 = \pm 4 $ , $c_{max}=1$ which is represented by a black dot where the curves for $p_2 = \pm 4 $ cross $-\alpha_{\max}$. For $p_2 = \pm 1, \pm 1$, $K^*(c,\bm{\theta})$ reaches a constant below $-\alpha_{\max}$ as $c$ grows, indicating that the basin of attraction has infinite radius.
Figure \ref{fig:cont:example}(\emph{B}) considers the case that $p_2= \pm 1$. We see two different regions in the $r_1(0),r_2(0)$-plane distinguished by two colors: the red region indicates the initial conditions from which the system's time evolution approaches the origin as time grows and the yellow region indicates the initial conditions from which the system's time evolution does not converge to the origin, in which case the dynamics converges to either another fixed point, or a limit cycle, or any other attractor other than the origin. The black circle  is the solution of $\|\textbf{r}\| = c_{\max}$, for $p_2= \pm 4$.  In Figs.\ \ref{fig:cont:example}(\emph{C}) and \ref{fig:cont:example}(\emph{D}) we plot the trajectory $r_2(t)$ versus $r_1(t)$ when the system is evolved from a typical initial condition from within the red and the yellow region, respectively.

\subsection{Reservoir Computer with Discrete Time Dynamics}

We now turn to the dynamics of a reservoir computer with discrete time dynamics,
\begin{equation}\label{eq:dis:dyn1}
{r}_i(n+1) =  f(r_i(n),\bm{\theta}) + \sum_{j = 1}^{M} A_{ij} r_j(n) + w_i s(n), \quad i = 1,2,\cdots, M,
\end{equation}
which in the unforced case becomes, 
\begin{equation}\label{eq:dis:dyn2}
{r}_i(n+1) =  f(r_i(n),\bm{\theta}) + \sum_{j = 1}^{M} A_{ij} r_j(n), \quad i = 1,2,\cdots, M,
\end{equation}
where $r_i(n) \in \mathbb{R}$, $i = 1,2,\cdots,M$, denotes the state of the  node $i$ of the reservoir at time step $n$, $f(r_i,\bm{\theta})$ is the nonlinear nodal dynamics of node $i$, the adjacency matrix $A=\{A_{ij}\}$ indicates the pattern of connectivity between the network nodes, $s(n) \in \mathbb{R}$ is the input signal at time step $n$ and $\textbf{w} = \left[w_1 , w_2,\cdots,w_M \right]$ is a vector that describes the coupling of the input signal $s(n)$ to each one of the nodes. The input signal $s(n)$ in Eq.\ \eqref{eq:dis:dyn1} is normalized to have mean $0$ and standard deviation equal to $1$ \cite{lu2017reservoir,carroll2019network}.

Hereafter we assume that the operating fixed point for the dynamics Eq.\ \eqref{eq:dis:dyn2} coincides with the origin (however, this assumption can be removed; see the example that follows for the case of a sigmoid nonlinearity.) %In case it coincides with a point different from the origin, we will use a coordinate transformation which moves this fixed point to the origin (see the example that follows for the case of a sigmoid nonlinearity.)

\subsubsection{Lyapunov Function and $c(\bm{\theta})$-region Design}\label{subsubsec:dis:lyap}

\begin{figure}
	\centering
	\includegraphics[width=1\textwidth]{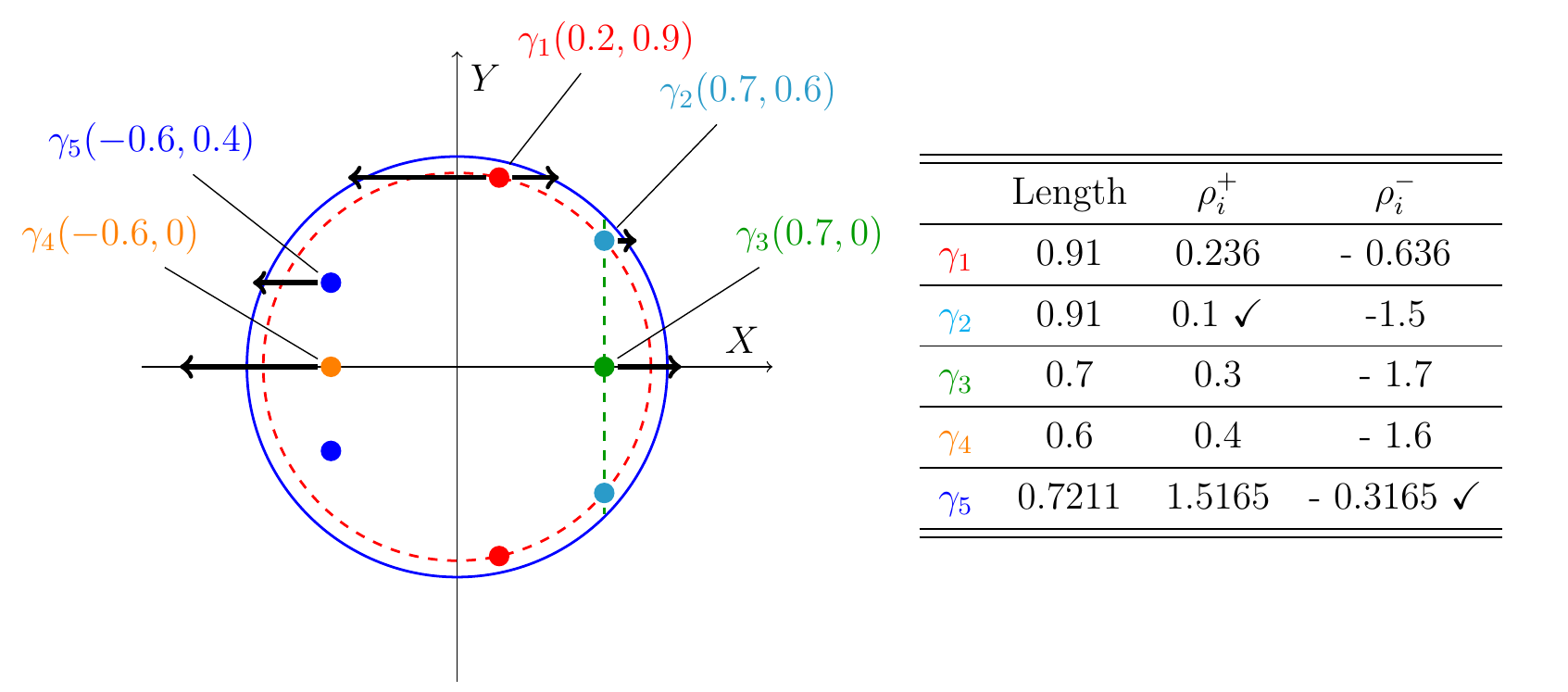}
	\caption{\textbf{Illustration of the critical eigenvalues appeared in the derivation of the scalar functions $K^{-*}(c,\bm{\theta})$ and  $K^{+*}(c,\bm{\theta})$ for the case of discrete time dynamics.} An example of several eigenvalues of some hypothetical adjacency matrix $A$ are shown inside the unit circle. The table lists for each eigenvalue the length or magnitude, the  shift $\rho_i^+$ to the positive $X$-direction, and the  shift $\rho_i^-$ to the negative $X$-direction. The critical eigenvalue closest to the positive side of the unit circle Wis $\gamma_2$ and the critical eigenvalue closest to the negative side of the unit circle is $\gamma_5$ when moving only along the real axis.
	}
	\label{fig:dis:example}
\end{figure}

We define a Lyapunov function $V : \mathcal{G}(c) \subseteq \mathbb{R}^M \rightarrow \mathbb{R}$ 
\begin{equation}\label{eq:dis:lyap}
V(\textbf{r}) = \| \textbf{r}\|
\end{equation}
where $\mathcal{G}(c) = \left\{ \textbf{r} = (r_1,r_2,\cdots,r_M ) \in \mathbb{R}^M :  \|\textbf{r}\| \le c \right\}$ and $\|\textbf{r}\| = \sqrt{r_1^2+\cdots+r_M^2}$. Here $V(\textbf{r})>0$ for $\textbf{r} \ne 0$ and $V(\textbf{0})= 0$. Then $\text{ for all }  \textbf{r} \in \mathcal{G}(c)\backslash \textbf{0} $, 
\begin{subequations}
	\begin{align}
	V(f(\textbf{r},\bm{\theta})+A\textbf{r}) - V(\textbf{r}) \\
	=	\|f(\textbf{r},\bm{\theta}) +A\textbf{r}\| - \|\textbf{r}\|  
	\label{eq:dis:eq:cont:lyap1}
	\end{align}
\end{subequations}
%
%where $\bar{\gamma} = \bar{\gamma}^{re}+j \bar{\gamma}^{im} $ is an eigenvalue of the matrix $A$ for which $\|f(\textbf{r},\bm{\theta}) +A\textbf{r}\| - \|\textbf{r}\|  
%	\le \|f(\textbf{r},\bm{\theta}) + \bar{\gamma}\textbf{r}\| - \|\textbf{r}\| $.
% 
We seek to find a scalar function $K(c,\bm{\theta})$ such that $f(r_i,\bm{\theta}) \le K(c,\bm{\theta}) r_i$ which also satisfies the inequality in Eq.\ \eqref{eq:dis:eq:cont:lyap1},
\begin{equation}\label{eq:dis:ineq1}
\|f(\textbf{r},\bm{\theta}) + A\textbf{r}\| - \|\textbf{r}\| \le
\|K(c,\bm{\theta}) \textbf{r}  + A \textbf{r}\| - \|\textbf{r}\|
\end{equation}
According to the Lyapunov stability theorem for discrete time dynamics \cite{haddad2011nonlinear}, the system is stable only if,
\begin{subequations}
	\begin{align}
	& |K(c,\bm{\theta})I  + A|  \|\textbf{r}\| - \|\textbf{r}\| \le 0 &&\\
	\text{or equivalently, } \quad  & |K(c,\bm{\theta}) +  \gamma_i| \le 1, && \text{ for } i = 1,2,\cdots,M
	\end{align}
\end{subequations}
where $\gamma_i = \gamma_i^{re}+j \gamma_i^{im} $ is an eigenvalue of the matrix $A$ and $j = \sqrt{-1}$.
The above inequality can be written as,
\begin{subequations}
	\begin{align}
	|K + \gamma_i^{re}| \le \sqrt{1 - (\gamma_i^{im})^2}\\
	- \sqrt{1 - (\gamma_i^{im})^2} \le K + \gamma_i^{re} \le  \sqrt{1 - (\gamma_i^{im})^2}
	\end{align}
\end{subequations}
We see that there is both an upper bound and a lower bound for $K(c,\bm{\theta})$.
%We see that $|K(c,\bm{\theta})+\gamma_i^{re}|$ can exceed $ \sqrt{1 - (\gamma_i^{im})^2}$ in two different ways: one is the positive $X$-direction and the other one is the negative $X$-direction (see Fig.\ 2). 
%
Hence, there are two critical eigenvalues: $\gamma_{c+}$ and $\gamma_{c-}$ which are the eigenvalues closest to the positive side of the unit circle and closest to the negative side of the unit circle when moving only along the real axis, respectively.
This concept is displayed graphically in Fig. \ref{fig:dis:example} where $\gamma_2 = \gamma_{c+}$ and $\gamma_5 = \gamma_{c-}$.
%Hence, there are two critical eigenvalues: $\gamma_{c+}$ on the positive side of the imaginary axis and  $\gamma_{c-}$ on the negative side of of the imaginary axis. 
The maximum distance the eigenvalue $\gamma_{c+}$ can shift to the right is
$\rho_{c+}^+ =   \sqrt{1 - (\gamma_{c+}^{im})^2} - \gamma_{c+}^{re} $ and the maximum distance the eigenvalue $\gamma_{c-}$ can shift to the left is
$\rho_{c-}^- =  -( \sqrt{1 - (\gamma_{c-}^{im})^2} + \gamma_{c-}^{re}) $.
Thus there exist two scalar functions denoted by $K(c,\bm{\theta}) = K^-(c,\bm{\theta}) \le 0  $ and $K(c,\bm{\theta}) = K^+(c,\bm{\theta})\ge 0 $ such that,
\begin{equation}\label{eq:dis:ineq2}
\begin{aligned}
\rho_{c-} ^-   \le  K^-(c,\bm{\theta}) \le  K^+(c,\bm{\theta})  \le \rho_{c+}^+  
\end{aligned}
\end{equation}
An illustration is presented in Fig.\ \ref{fig:dis:example} which shows how to find the critical eigenvalues $\gamma_{c+}$ and $\gamma_{c-}$. 
In Fig.\ \ref{fig:dis:example}, several eigenvalues of some hypothetical adjacency matrix $A$ are shown inside the unit circle.  For each eigenvalue, we compute $\rho_i^+$ and $\rho_i^-$. From the table we see that $\gamma_2$ is the critical eigenvalue $\gamma_{c+}$ and   $\gamma_5$ is the critical eigenvalue $\gamma_{c-}$.

Now using the fact that all the nodes are homogeneous, from inequality  Eq.\ \eqref{eq:dis:ineq1}, we can write,
\begin{equation}\label{eq:dis:ineq3}
\begin{aligned}
0 \le \left| \frac{f(r_i,\bm{\theta}) }{r_i} + \gamma_i^{re} \right|& \le |K + \gamma_i^{re}| \le  \sqrt{1 - (\gamma_i^{im})^2}
\end{aligned}
\end{equation}
From the inequalities in Eqs.\ \eqref{eq:dis:ineq2} and \eqref{eq:dis:ineq3}, it follows that,
\begin{equation}
\begin{aligned}
%	\rho_{c-}^- + \bar{\gamma}^{re} \le  K^-(c,\bm{\theta}) + \bar{\gamma}^{re}   \le &\frac{f(r_i,\bm{\theta}) }{r_i} + \bar{\gamma}^{re}  \le K^+(c,\bm{\theta}) + \bar{\gamma}^{re}   \le K^+ + \bar{\gamma}^{re} \\
\rho_{c-}^-  \le  K^-(c,\bm{\theta})    \le &\frac{f(r_i,\bm{\theta}) }{r_i}   \le K^+(c,\bm{\theta})    \le \rho_{c+}^+  \\
\end{aligned}
\end{equation}
Thus, we find $K^-(c,\bm{\theta})$ and $K^+(c,\bm{\theta})$ such that 
\begin{subequations}
	\begin{align}
	\rho_{c-}^-  \le  K^-(c,\bm{\theta})  \le \frac{f(r_i,\bm{\theta}) }{r_i}  \\
	\text{and } \quad	 \frac{f(r_i,\bm{\theta}) }{r_i}  \le K^+(c,\bm{\theta}) \le  \rho_{c+}^+
	\end{align}
\end{subequations}
As we want tight upper and lower bounds on $\frac{f(r_i,\bm{\theta}) }{r_i} $, we seek to find $K^+(c,\bm{\theta})$ and $K^-(c,\bm{\theta})$  that solve the following two optimization problems,
\begin{equation}\label{eq:dis:opt1}
\begin{aligned}
\max & &&  K^-(c,\bm{\theta}) \\
s.t. & && f(r_i,\bm{\theta}) - K^-(c,\bm{\theta}) r_i   \ge  0, \quad \forall r_i \in [-c,c]
\end{aligned}
\end{equation} 
and
\begin{equation}\label{eq:dis:opt2}
\begin{aligned}
\min & &&K^+(c,\bm{\theta}) \\
s.t. & && f(r_i,\bm{\theta}) - K^+(c,\bm{\theta})r_i   \le 0, \quad \forall r_i \in [-c,c]
\end{aligned}
\end{equation} 
A solution ${K^-}^*(c,\bm{\theta})$ to the problem in Eq.\ \eqref{eq:dis:opt1} and  ${K^+}^*(c,\bm{\theta})$ to the problem in Eq.\ \eqref{eq:dis:opt2} must satisfy their respective constraints exactly for some $r_i^{+*}$ and $r_i^{-*}$,
\begin{subequations}
	\begin{align}
	{K^-}^*(c,\bm{\theta})r_i^{-*} - f(r^{-*}_i,\bm{\theta}) & = 0, \quad r_i^{-*} \in [-c,c] \\
	\text{and } {K^+}^*(c,\bm{\theta})r_i^{+*} - f(r^{+*}_i,\bm{\theta}) & = 0, \quad r_i^{+*} \in [-c,c] 
	\end{align}
\end{subequations} 
or equivalently,
\begin{subequations}
	\begin{align}
	{K^-}^*(c,\bm{\theta}) & = \frac{f(r^{-*}_i,\bm{\theta})}{r^{-*}_i} \le  \frac{f(r_i,\bm{\theta})}{r_i} \quad \forall r_i \in [-c,c]\\
	\text{and } 	{K^+}^*(c,\bm{\theta}) & = \frac{f(r^{+*}_i,\bm{\theta})}{r^{+*}_i} \ge  \frac{f(r_i,\bm{\theta})}{r_i} \quad \forall r_i \in [-c,c]
	\end{align}
\end{subequations} 
We can find ${K^-}^*(c,\bm{\theta})$ as
\begin{equation}\label{eq:dis:kstar1}
{K^-}^*(c,\bm{\theta}) = \min 		
\begin{cases}
\frac{f(c,\bm{\theta})}{c},\\
\frac{f(-c,\bm{\theta})}{-c},\\
f'(0,\bm{\theta}),\\
\frac{f(r^*_i,\bm{\theta})}{r_i^*}, \text{ where } r^*_i \text{ is the root of }\\
\quad \quad \quad r^*_if'(r^*_i,\bm{\theta})-f(r^*_i,\bm{\theta}) = 0,  r^*_i \in[-c,c], r^*_i \ne 0 
\end{cases}
\end{equation} 
and we can find ${K^+}^*(c,\bm{\theta})$ as
\begin{equation}\label{eq:dis:kstar2}
{K^+}^*(c,\bm{\theta}) = \max		
\begin{cases}
\frac{f(c,\bm{\theta})}{c},\\
\frac{f(-c,\bm{\theta})}{-c},\\
f'(0,\bm{\theta}),\\
\frac{f(r^*_i,\bm{\theta})}{r_i^*}, \text{ where } r^*_i \text{ is the root of }\\
\quad \quad \quad r^*_if'(r^*_i,\bm{\theta})-f(r^*_i,\bm{\theta}) = 0,  r^*_i \in[-c,c], r^*_i \ne 0 
\end{cases}
\end{equation}
Once we obtain ${K^-}^*(c,\bm{\theta})$ and ${K^+}^*(c,\bm{\theta})$, we can find $c_{\max}^+$ and $c_{\max}^-$ such that ${K^-}^*(c_{\max}^-,\bm{\theta}) = \rho_{c-}^-$ and ${K^+}^*(c_{\max}^+,\bm{\theta}) = \rho_{c+}^+$, respectively. The $c(\bm{\theta})$-region for the discrete time RC can be determined as $\mathcal{G}(c_{\max}) = \{ \textbf{r} = (r_1,r_2,\cdots,r_M) \in \mathbb{R}^M : \|\textbf{r}\| \le c_{\max} \}$, where $c_{\max}= \min\{c_{\max}^-, c_{\max}^+\}$. 

\subsubsection*{Example: $\tanh()$ type nonlinearity }

We choose the nodal dynamics to be
\begin{equation}
f(r_i,\bm{\theta}) = p_1\tanh(p_2 r_i), \quad p_2 > 0
\end{equation}
The dynamics of node $i$ is described by,
\begin{equation}\label{eq:dis:tanh}
r_i(n+1) = p_1\tanh(p_2 r_i(n))+ \sum_j A_{ij} r_j(n)+ w_i s(n)
\end{equation}
Here $\bm{\theta} = \{p_1, p_2\}$. We see that for $s(n)=0$, the origin is a fixed point, which is stable if all the eigenvalues of the matrix $(A +p_1 p_2 I)$ are inside the unit circle.
The constant functions ${K^-}^*(c,\bm{\theta})$ and ${K^+}^*(c,\bm{\theta})$ can be found as,
\begin{equation}
\begin{aligned}
{K^-}^*(c,\bm{\theta}) & =  \min		
\begin{cases}
\frac{f(c,\bm{\theta})}{c} = \frac{p_1\tanh(p_2c)}{c},  \\
\frac{f(-c,\bm{\theta})}{-c} = \frac{p_1\tanh(p_2c)}{c}, \\
f'(0,\bm{\theta}) = p_1 p_2     
\end{cases} &&& &&&  &&  &&&&& &&&&& \\
\end{aligned}
\end{equation} 
and 
\begin{equation}
\begin{aligned}
{K^+}^*(c,\bm{\theta}) & =  \max		
\begin{cases}
\frac{f(c,\bm{\theta})}{c} = \frac{p_1\tanh(p_2c)}{c},  \\
\frac{f(-c,\bm{\theta})}{-c} = \frac{p_1\tanh(p_2c)}{c}, \\
f'(0,\bm{\theta}) = p_1      p_2
\end{cases} &&& &&&  &&  &&&&& &&&&& \\
\end{aligned}
\end{equation}
%=
Now if $p_1 < 0$, then $p_1 p_2 = {K^-}^*(c,\bm{\theta}) \le 0$ and if $p_1>0$, then $ 0 \le {K^+}^*(c,\bm{\theta}) = p_1p_2$ for any choice of $c$.

\subsubsection{Lyapunov Function and $c(\bm{\theta})$-region Design for Non-homogeneous Nodal Dynamics}

One generalization of Eq.\ \eqref{eq:dis:dyn2} is to the case of non-homogeneous nodal dynamics,
\begin{equation} \label{eq:nonh}
\bar{r}_i(n+1) =  \bar{f}_i(\bar{r}_i(n),\bm{\theta}_i)  + \sum_j A_{ij} \bar{r}_j(n).
\end{equation} 
Without loss of generality we retain the assumption that the above set of equations has a fixed point at the origin. In the case a fixed point exists that is different from the origin, this assumption can be removed by applying a coordinate transformation that moves the fixed point to the origin (see the example that follows for the case of a sigmoid nonlinearity). Now according to the Lyapunov function analysis described in section B.1, for each node $i$ we find scalar functions $K_i^{-*}(c,\bm{\theta}_i)$ and $K_i^{+*}(c,\bm{\theta}_i)$ which satisfy,
\begin{subequations}
	\begin{align}
	{K_i^-}^*(c,\bm{\theta}_i) & \le  \frac{\bar{f}_i(\bar{r}_i,\bm{\theta}_i)}{\bar{r}_i} \text{ over the interval } \bar{r}_i \in [-c,c]\\
	\text{and } 	{K_i^+}^*(c,\bm{\theta_i}) &  \ge  \frac{\bar{f}_i(\bar{r}_i,\bm{\theta}_i)}{\bar{r}_i} \text{ over the interval } \bar{r}_i \in [-c,c]
	\end{align}
\end{subequations} 
We can find ${K_i^-}^*(c,\bm{\theta}_i)$ as
\begin{equation}\label{eq:dis:non-hom:Km}
{K_i^-}^*(c,\bm{\theta}_i) = \min 		
\begin{cases}
\frac{\bar{f}_i(c,\bm{\theta}_i)}{c},\\
\frac{\bar{f}_i(-c,\bm{\theta}_i)}{-c},\\
\bar{f}_i'(0,\bm{\theta}_i),\\
\frac{\bar{f}_i(\bar{r}^*_i,\bm{\theta}_i)}{\bar{r}_i^*}, \text{ where } \bar{r}^*_i \text{ is the root of }\\
\quad \quad \quad \bar{r}^*_i \bar{f}_i'(\bar{r}^*_i,\bm{\theta}_i)-\bar{f}_i(\bar{r}^*_i,\bm{\theta}_i) = 0,  \bar{r}^*_i \in[-c,c], \bar{r}^*_i \ne 0 
\end{cases}
\end{equation} 
and we can find ${K_i^+}^*(c,\bm{\theta}_i)$ as
\begin{equation}\label{eq:dis:non-hom:Kp}
{K_i^+}^*(c,\bm{\theta}_i) = \max		
\begin{cases}
\frac{\bar{f}_i(c,\bm{\theta}_i)}{c},\\
\frac{\bar{f}_i(-c,\bm{\theta}_i)}{-c},\\
\bar{f}_i'(0,\bm{\theta}_i),\\
\frac{\bar{f}_i(\bar{r}^*_i,\bm{\theta}_i)}{\bar{r}_i^*}, \text{ where } \bar{r}^*_i \text{ is the root of }\\
\quad \quad \quad \bar{r}^*_i \bar{f}_i'(\bar{r}^*_i,\bm{\theta}_i)-\bar{f}_i(\bar{r}^*_i,\bm{\theta}_i) = 0,  \bar{r}^*_i \in[-c,c], \bar{r}^*_i \ne 0 
\end{cases}
\end{equation}
Now we define the scalar function ${K^+}^*(c,\bm{\theta})$ as,
\begin{equation}
{K^-}^*(c,\bm{\theta}) = \min_i  \left\{{K_i^-}^*(c,\bm{\theta}_i) \right\}
\end{equation}
and the scalar function ${K^+}^*(c,\bm{\theta})$ as,
\begin{equation}
{K^+}^*(c,\bm{\theta}) = \max_i  \left\{ {K_i^+}^*(c,\bm{\theta}_i) \right\}
\end{equation}
Once we obtain ${K^-}^*(c,\bm{\theta})$ and ${K^+}^*(c,\bm{\theta})$, we can find $c_{\max}^+$ and $c_{\max}^-$ such that ${K^-}^*(c_{\max}^-,\bm{\theta}) = \rho_{c-}^-$ and ${K^+}^*(c_{\max}^+,\bm{\theta}) = \rho_{c+}^+$, respectively. 
Then the $c$-region for the  reservoir computer can be determined as $\mathcal{G}(c_{\max}) = \{ \textbf{r} - \textbf{q}^* : \|\textbf{r} - \textbf{q}^*\| \le c_{\max} \}$, where $c_{\max} = \min\{c_{\max}^+,c_{\max}^-\}$.

\begin{comment}

Consider the following unforced dynamics,
\begin{equation}
r_i(n+1) = F(r_i(n)) = f(r_i(n),\theta) + \sum_j A_{ij} r_j(n),
\end{equation}
%
for which $q_i^*$, $i=1,...,M$ is a nonzero fixed point. Then,
%
\begin{equation}
F(q_i^*) = f(q_i^*,\bm{\theta}) + \sum_j A_{ij} q_j^* =  q_i^*
\end{equation}
%
Let us consider the following coordinate transformation
%
\begin{equation}
\bar{r}_i(n) = r_i(n) - q_i^*
\end{equation}
%
Then
\begin{equation}
F(\bar{r}_i(n) + q_i^*) = \bar{r}_i(n+1) + q_i^*
\end{equation}
%
If we set $ \bar{F}(\bar{r}_i(n)) = F(\bar{r}_i(n) + q_i^*) - q_i^*$, then 
%
\begin{equation}
\bar{r}_i(n+1)  = \bar{F}(\bar{r}_i(n)). 
\end{equation}
%
Now
\begin{subequations}
\begin{align}
\bar{r}_i(n+1) 
& = \bar{F}(\bar{r}_i(n)) \\
& = F(\bar{r}_i(n) + q_i^*) - q_i^*\\
& = f((\bar{r}_i(n) + q_i^*,\bm{\theta}) + \sum_j A_{ij} \bar{r}_j(n) +\sum_j A_{ij} q_j^* - q_i^*\\
& = f((\bar{r}_i(n) + q_i^*,\bm{\theta})+\sum_j A_{ij} q_j^* - q_i^* + \sum_j A_{ij} \bar{r}_j(n)
\end{align}
\end{subequations}
%
Let $f((\bar{r}_i(n) + q_i^*,\bm{\theta})+\sum_j A_{ij} q_j^* - q_i^* = \bar{f}_i(\bar{r}_i(n),\bm{\theta}_i)$ and $\bar{f}_i(\bar{r}_i(n),\phi_i) \rightarrow 0$ as $\bar{r}_i(n) \rightarrow 0$. Note that here the nodal dynamics $\bar{f}_i(\bar{r}_i(n),\bm{\theta}_i$ is not homogeneous because the term $\sum_j A_{ij} q_j^*$ can be different for each node $i$ and we have reparameterized the nodal dynamics.  
%
The transformed system can be written as,
\end{comment}

\subsubsection*{Example: Sigmoid Nonlinearity }

Here we choose the nodal dynamics to be
\begin{equation}
f(r_i,\bm{\theta}) = \frac{p_1}{1+e^{-p_2r_i}}
\end{equation}
The unforced dynamics of each node $i$ is,
\begin{equation}\label{eq:dis:sigmoid}
r_i(n+1) = \frac{p_1}{1+e^{-p_2r_i}} + \sum_j A_{ij} r_j(n)
\end{equation}
Here the parameters are $\bm{\theta} = \{p_1, p_2\}$. For this example, we see that the origin is not a fixed point. Instead a nonzero fixed point exists: $q_i^*$, $i=1,...,M$. Let $\bar{r}_i = r_i - q_i^*$ and the system in Eq.\ \eqref{eq:dis:sigmoid} can be transformed to the form of Eqs.\ \eqref{eq:nonh},
%
%\begin{equation} \label{eq:rbar}
%\bar{r}_i(n+1) = \bar{f}_i(\bar{r}_i(n),\bm{\theta}_i)  + \sum_j A_{ij} \bar{r}_j(n)
%\end{equation}
%
where $\bar{f}_i(\bar{r}_i(n),\bm{\theta}_i) = f((\bar{r}_i(n) + q_i^*,\bm{\theta})+\sum_j A_{ij} q_j^* - q_i^*$. %Eqs.\eqref{eq:rbar} is the same as Eqs.\ \eqref{eq:nonh}.  %and $q_i^*$ is a fixed point of $F(r_i(n)) = \frac{p_1}{1+e^{-p_2r_i}} + \sum_j A_{ij} r_j(n)$. 
According to Eq.\ \eqref{eq:dis:non-hom:Km} and \eqref{eq:dis:non-hom:Kp}, we find the scalar functions   ${K_i^-}^*(c,\bm{\theta}_i)$ and ${K_i^+}^*(c,\bm{\theta}_i)$ as follows,
\begin{equation}\label{eq:dis:non-hom:Km:sigmoid}
{K_i^-}^*(c,\bm{\theta}_i) = \min 		
\begin{cases}
\frac{\bar{f}_i(c,\bm{\theta_i})}{c} = \frac{\frac{p_1}{1+e^{-p_2( c+q_i^*)}} + \sum_j A_{ij} q_j^* - q_i^* }{c},\\
\frac{\bar{f}_i(-c,\bm{\theta_i})}{-c} =  \frac{\frac{p_1}{1+e^{-p_2( -c+q_i^*)}} + \sum_j A_{ij} q_j^* - q_i^* }{-c},\\
\bar{f}_i'(0,\bm{\theta}_i) = \frac{p_1p_2e^{-p_2q_i^*}}{(1+e^{-p_2q_i^*})^2}
\end{cases}
\end{equation} 
and
\begin{equation}\label{eq:dis:non-hom:Kp:sigmoid}
{K_i^+}^*(c,\bm{\theta}_i) = \max		
\begin{cases}
\frac{\bar{f}_i(c,\bm{\theta_i})}{c} = \frac{\frac{p_1}{1+e^{-p_2( c+q_i^*)}} + \sum_j A_{ij} q_i^* - q_i^* }{c},\\
\frac{\bar{f}_i(-c,\bm{\theta_i})}{-c} =  \frac{\frac{p_1}{1+e^{-p_2( -c+q_i^*)}} + \sum_j A_{ij} q_i^* - q_i^* }{-c},\\
\bar{f}_i'(0,\bm{\theta}_i) = \frac{p_1p_2e^{-p_2q_i^*}}{(1+e^{-p_2q_i^*})^2}
\end{cases}
\end{equation}
%=
Now we define the scalar function ${K^-}^*(c,\bm{\theta})$ as follows,
\begin{equation}\label{eq:dis:non-hom:Km:sigmoid2}
{K^-}^*(c,\bm{\theta}) = \min_i  \left\{{K_i^-}^*(c,\bm{\theta}_i) \right\}
\end{equation}
and the scalar function ${K^+}^*(c,\bm{\theta})$ as follows,
\begin{equation}\label{eq:dis:non-hom:Kp:sigmoid2}
{K^+}^*(c,\bm{\theta}) = \max_i  \left\{ {K_i^+}^*(c,\bm{\theta}_i) \right\}
\end{equation}
Once we obtain ${K^-}^*(c,\bm{\theta})$ and ${K^+}^*(c,\bm{\theta})$, we can find $c_{\max}^+$ and $c_{\max}^-$ such that ${K^-}^*(c_{\max}^-,\bm{\theta}) = \rho_{c-}^-$ and ${K^+}^*(c_{\max}^+,\bm{\theta}) = \rho_{c+}^+$, respectively. 
Then the $c$-region for the  reservoir computer can be determined as $\mathcal{G}(c_{\max}) = \{ \textbf{r} - \textbf{q}^* : \|\textbf{r} - \textbf{q}^*\| \le c_{\max} \}$, where $c_{\max} = \min\{c_{\max}^+,c_{\max}^-\}$.

\section{\label{sec:level2}Results}
\noindent
For our numerical simulations, in both continuous-time and discrete-time, we construct the adjacency matrix $A$ as follows: (i) We set the entries of the initial matrix $A$ to be equal to $A_{ij}=1-\delta_{ij}$, where $\delta_{ij}$ is the Kronecker delta, $i,j=1,...,M$. (ii)  $50 \%$ of the off-diagonal entries of the matrix $A$ are chosen randomly and set to zero. (iii)  $50 \%$ of the remaining nonzero entries of the matrix $A$ are chosen randomly and are flipped from $+1$ to $-1$. (iv) Finally, the adjacency matrix $A$ is normalized so that the absolute value of the largest real part of its eigenvalues is equal to $0.5$. 

\subsection{Training Error of a Reservoir Computer}

The training error, $\Delta_{RC}$, is used to quantify how well the training signal $g(t)$ ($g(n)$, in the case of discrete dynamics) can be reconstructed from the input signal $s(t)$ ($s(n)$). Lower values of  $\Delta_{RC}$ indicate a better performance of the reservoir computer. In the continuous-time case, the training signal and the input signal are discretized in time and are thus treated as sequences.
Before driving the reservoir computer by the input signal $s(t)$ ($s(n)$), both $s(t)$ and $g(t)$ ($s(n)$ and $g(n)$) are normalized to have mean equal to zero and standard deviation equal to one \cite{lu2017reservoir}. \FS{However, such a choice is completely arbitrary; the amplitude of the input signal can be conveniently reduced in case one finds the (stable) reservoir dynamics to become unstable when driven by the input signal.} Next, we present the procedure to compute the training error in the case of discrete-time dynamics (the procedure for the case of continuous-time dynamics is analogous.)
We set the number of nodes of the reservoir to $M=100$. When the reservoir  is driven by the input signal $s(n)$, the first 2000 time steps are discarded as a transient. The next $N = 10,000$ time steps from each node are recorded and combined into the $N \times (M+1)$ matrix, 
\begin{equation}
\Omega = \begin{bmatrix}
r_1(1) &	\cdots & r_M(1)	& 1\\
r_1(2) &	\cdots & r_M(2)	& 1\\
\vdots &	\vdots & \vdots	& 1\\
r_1(N) &	\cdots & r_M(N)	& 1\\
\end{bmatrix}
\end{equation}
The last column of the matrix $\Omega$ is set to 1 to account for any constant offset in the fitting. We then introduce
\begin{equation}
{\textbf h} = \Omega \textbf{k}
\end{equation}
where ${\textbf h} = \left[ h(1), h(2), \cdots, h(N)\right]$ is the fit to the training signal ${\textbf g}= \left[ g(1), g(2), \cdots, g(N)\right]$ and $\textbf{k} = \left[ k_1, k_2, \cdots, k_{M+1}\right]^T$  is the vector of coefficients. The  vector $\textbf{k}$ is obtained  from the minimum-norm solution to the linear least squares problem,
\begin{equation}
\min \quad ||\Omega \textbf{k} -{\textbf g}||
\end{equation}
The training error is computed as, 
\begin{equation}
\Delta_{RC} = \frac{\langle \Omega \textbf{k} - {\textbf g} \rangle}{ \langle {\textbf g} \rangle},
\end{equation}
where the symbol $\langle \cdot \rangle$ is computed by using the following formula,
\begin{equation}
\langle{ \textbf{X} \rangle} = \sqrt{\frac{1}{N} \sum_{i=1}^N \left(X(i)-\mu\right)^2}
\end{equation}
where $\textbf{X} = \left[ X(1), X(2), \cdots, X(N) \right]$ and $\mu = \frac{1}{N} \sum_{i=1}^N X(i)$.
%\textcolor{red}{Can you please add a formula for what $\langle.\rangle$ exactly means. I don't like `standard deviation'}

\subsection{Results for Continuous-Time Dynamics}

We now consider continuous-time and use \FS{the general polynomial function for the individual nodal dynamics,
	\begin{equation} \label{general}
	f({r_i,\bm{\theta}})= \sum_i p_i r^i
	\end{equation}
	of which
	Eq.\  \eqref{eq:cont:poly} is an example}. We keep $p_1$ constant ($p_1 = -3$) as we are mainly interested in the effects of the nonlinear terms. \FS{In what follows we will %consider different types of polynomials  and 
	study the effects of varying different pairs of $p_i$ coefficients, $i>1$, of the polynomial \eqref{general} (in addition to setting $p_1=-3$.) For different cases we will indicate the pairs of coefficients we are focusing on, with the understanding that all the remaining coefficients are set to zero. We see that  carefully  choosing the form of the polynomial \eqref{general} is important and that setting certain coefficients $p_i$ to zero can dramatically worsen the performance of the reservoir, even when this is not directly predicted by the stability analysis.  }

\begin{figure}
	\centering
	\includegraphics[width=0.8\textwidth]{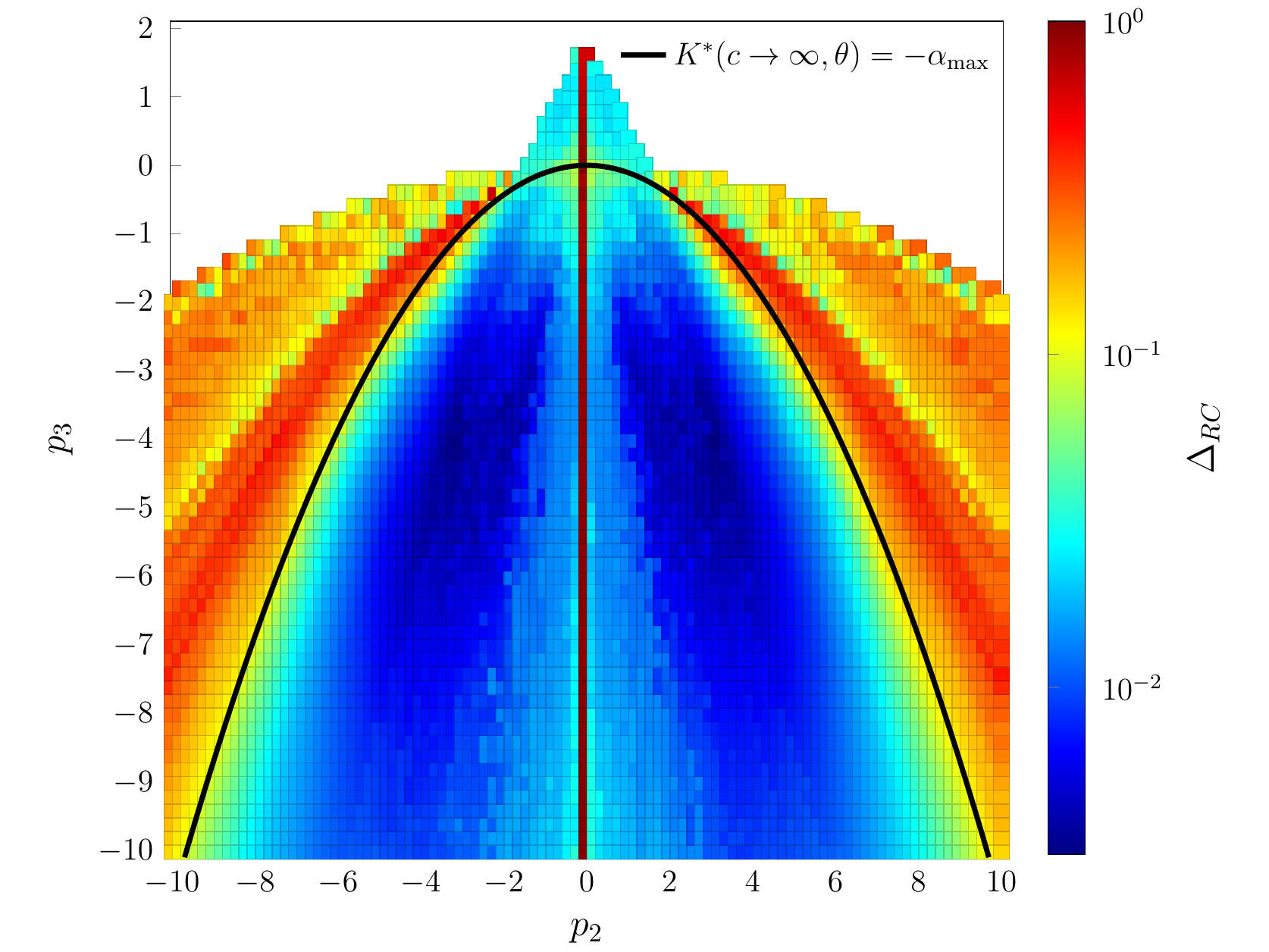}
	\caption{\textbf{Global stability and the RC performance for the case of continuous time and polynomial nodal dynamics. Input and training signals are from the Lorenz attractor.}  The training error $\Delta_{RC }$ is plotted on a log scale as the parameters $p_2$ and $p_3$ are varied.  The color varies from dark blue (small error) to dark red (large error).  The solid black curve represents for each value of $p_2$ the value of $p_3$ such that $K^*(c\rightarrow \infty,\bm{\theta}) = -\alpha_{\max}(A_s)$.}
	\label{fig:cont:poly1}
\end{figure}

\begin{figure}
	\centering
	\includegraphics[width=0.8\textwidth]{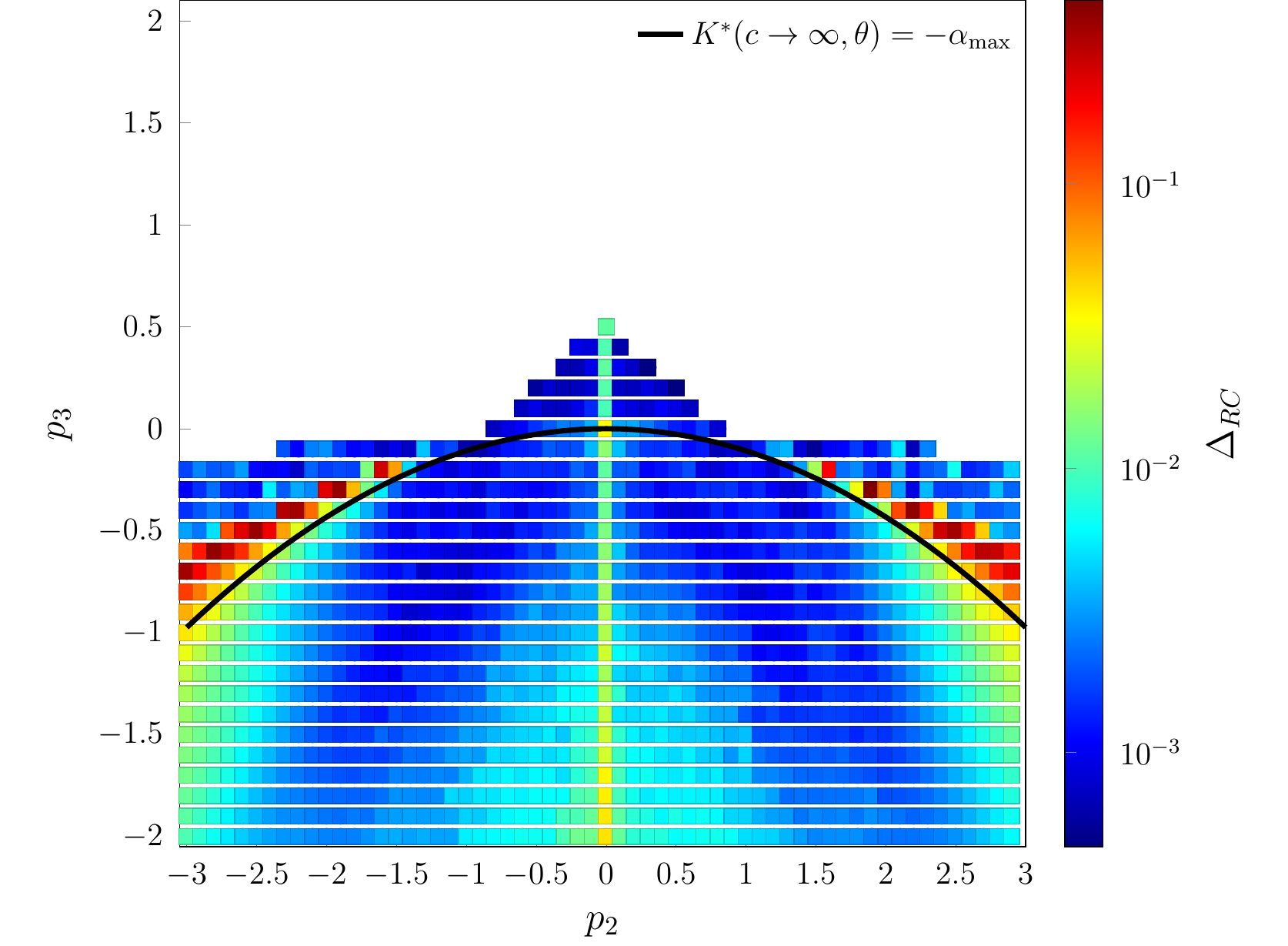}
	\caption{\textbf{Global stability and the RC performance for the case of continuous time and polynomial nodal dynamics. Input and training signals are from the Duffing attractor.} The training error $\Delta_{RC }$ is plotted on a log scale as the parameters $p_2$ and $p_3$ are varied.  The color varies from dark blue (small error) to dark red (large error).   The solid black curve represents for each value of $p_2$ the value of $p_3$ such that $K^*(c\rightarrow \infty,\bm{\theta}) = -\alpha_{\max}(A_s)$.}
	\label{fig:cont:poly2}
\end{figure}

\begin{figure}
	\centering
	\includegraphics[width=0.8\textwidth]{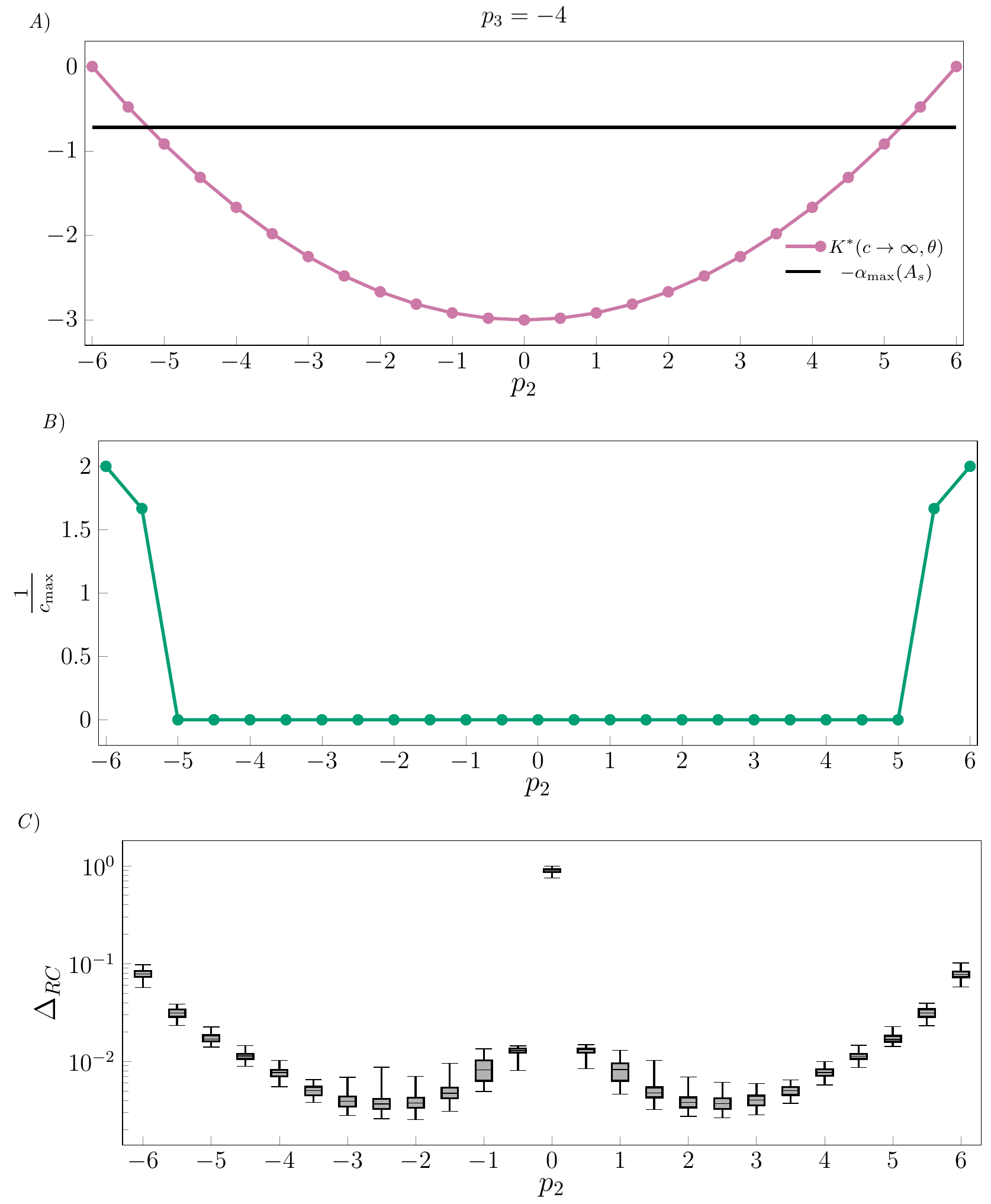}
	\caption{\textbf{Basin of attraction and the RC performance for the case of continuous time and polynomial nodal dynamics. Input and training signals are from the Lorenz attractor.} 
		(\emph{A}) ${K}^*(c,\bm{\theta})$ (magenta)   versus  $p_2$ for the case that $p_3 = -4$.
		(\emph{B}) The inverse of the radius of the $c$-region (${1}/{c_{\max}}$)  versus the parameter $p_2$ for the case that $p_3 = -4$.
		(\emph{C}) The training error ($\Delta_{RC}$)  versus the parameter $p_2$ for the case that $p_3 = -4$. }
	\label{fig:cont:poly3}
\end{figure}

Figure \ref{fig:cont:poly1}  provides a visual assessment of the training error of the  reservoir computer in terms of the parameters $p_2$ and $p_3$.
First we construct the matrix $A$ as described at the beginning of this section.
For the input signal $s(t)$ we choose the $x$ signal from a Lorenz chaotic attractor, while the training signal $g(t)$ is the Lorenz $z$ signal \cite{pathak2017using}.
The input signal $s(t)$  and the training signal $g(t)$ are normalized to have mean equal to zero and standard deviation equal to one.
In Fig.\ \ref{fig:cont:poly1} we plot the training error $\Delta_{RC }$  as a function of the parameters $p_2$ and $p_3$.
The color represents variations in log scale of  the training error from dark blue (small) to dark red (large).
The solid black curve represents the boundary between sets of parameters such that the RC is globally stable (below the black line) versus sets of parameters such that $c_{\max}$ is finite (above the black curve).
%$K^*(c,\bm{\theta}) = -\alpha_{\max}(A_s)$ as $c\rightarrow \infty$ as a function of $p_2$ and $p_3$.
In other words, the region under the black curve  determines the part of the parameter space $(p_2,p_3)$ for which the reservoir computer is globally stable  and the RC is successful in performing the computation (though the training error { may vary by different orders of magnitude as the parameters change.})
On the other hand, above the black curve, it is difficult to assess how the system behaves. For example, the system could be globally stable, or locally stable, in which case depending on the particular choice of the input, the system dynamics might approach a different attractor or might be driven to $\pm \infty$.  Another interesting observation is the presence of a tiny triangular region above the black curve where the reservoir computer performs well. %However, while our numerical results seem to indicate that this is a good region in the parameter space, it is still possible for a little perturbation in the parameters to drive the system to $\pm \infty$.
The white region is the area of the parameter space for which the system goes to $\pm \infty$  when it is driven by the input signal and the reservoir computer fails in performing the computation. For $p_3 > 2$ the RC dynamics diverges independent of the choice of $p_2$.
We also notice that the training error is symmetric about the $p_2 = 0$-axis and the training error is very high (almost equals to 1) at $p_2 = 0$, which indicates that the reservoir computer fails to capture and transfer the information from input to output if the quadratic term is absent from the nodal dynamics. \FS{Note that this seems to indicate there are two distinct requirements for the reservoir to work properly: (i) it needs to operate in the area of global stability and (ii) the $p_2$ coefficient needs to be nonzero.}
We observe a similar behavior in Fig.\ \ref{fig:cont:poly2} where the input signal $s(t)$ is the $x$- component and the training signal is the $y$-component of the Duffing chaotic attractor \cite{chang2017chaotic}. 
Figure\ \ref{fig:cont:poly3} shows in further detail the results in Fig.\ \ref{fig:cont:poly1} for $p_3 = -4$ while preserving the nodal dynamics, the adjacency matrix $A$, etc. 
Figure\ \ref{fig:cont:poly3}(\emph{A}) is a plot of $K^*(c \rightarrow \infty,\bm{\theta})$ versus the parameter $p_2$ when $p_3 = -4$ and the black line is the constant-ordinate line at $- \alpha_{\max}$ of the symmetric part of the matrix $A$.
In Fig.\ \ref{fig:cont:poly3}(\emph{B}), we plot the inverse of the radius of the $c(\bm{\theta})$-region ($\frac{1}{c_{\max}}$)  versus the parameter $p_2$ for the particular case of $p_3 = -4$. 
Here $\frac{1}{c_{\max}} = 0$ indicates that the system is globally stable and $\frac{1}{c_{\max}} = \infty $ indicates that the system is unstable. Intermediate values of $\frac{1}{c_{\max}}$ indicate that the system is $c$-radius stable.
In Fig.\ \ref{fig:cont:poly3}(\emph{C}), we present a box plot of the training error ($\Delta_{RC}$)  versus the parameter $p_2$  ($p_3 = -4$). In this simulation, we consider 100 different realizations of the matrix $A$. The training signal $s(t)$ is the $x$-component and the input signal $g(t)$ is the $z$-component of the Lorenz attractor. 

\begin{figure}
	\centering
	\includegraphics[width=0.8\textwidth]{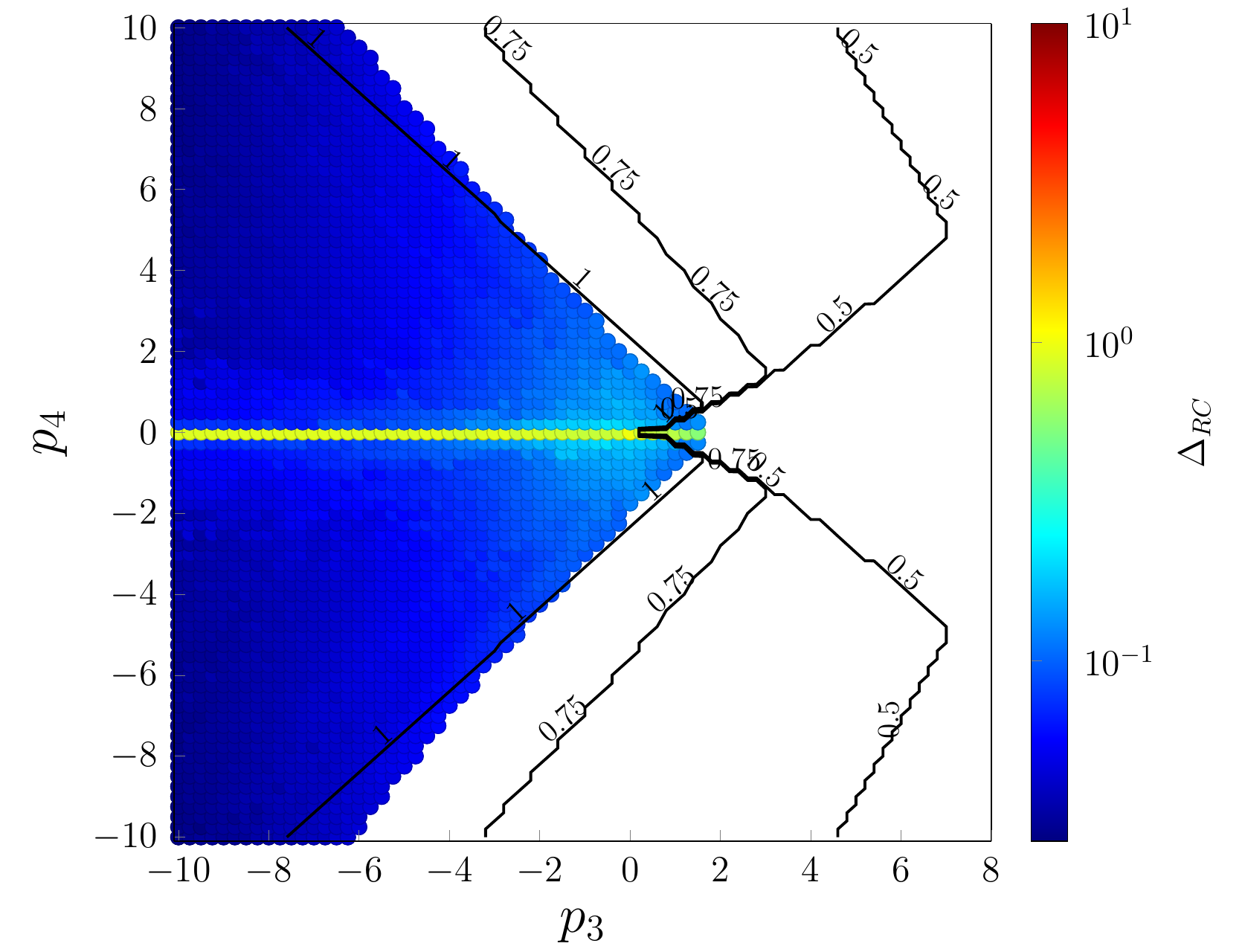} 
	\caption{\textbf{Finite region stability and the RC performance for the case of continuous time and polynomial nodal dynamics with $f(r,\boldsymbol{\theta}) = p_1 r +p_3 r^3 + p_4 r^4$. Input and training signals are from the Lorenz attractor.}  The training error $\Delta_{RC }$ is plotted on a log scale as the parameters $p_3$ and $p_4$ are varied.  The color varies from dark blue (small error) to dark red (large error).  The solid black curves represent the level curves for different values of $c_{\max}$ and $K^*(c_{\max},\bm{\theta}) = -\alpha_{\max}(A_s)$ in the parameter space $(p_3, p_4)$.}
	\label{fig:cont:poly4}
\end{figure}

\begin{figure}
	\centering
	\includegraphics[width=0.8\textwidth]{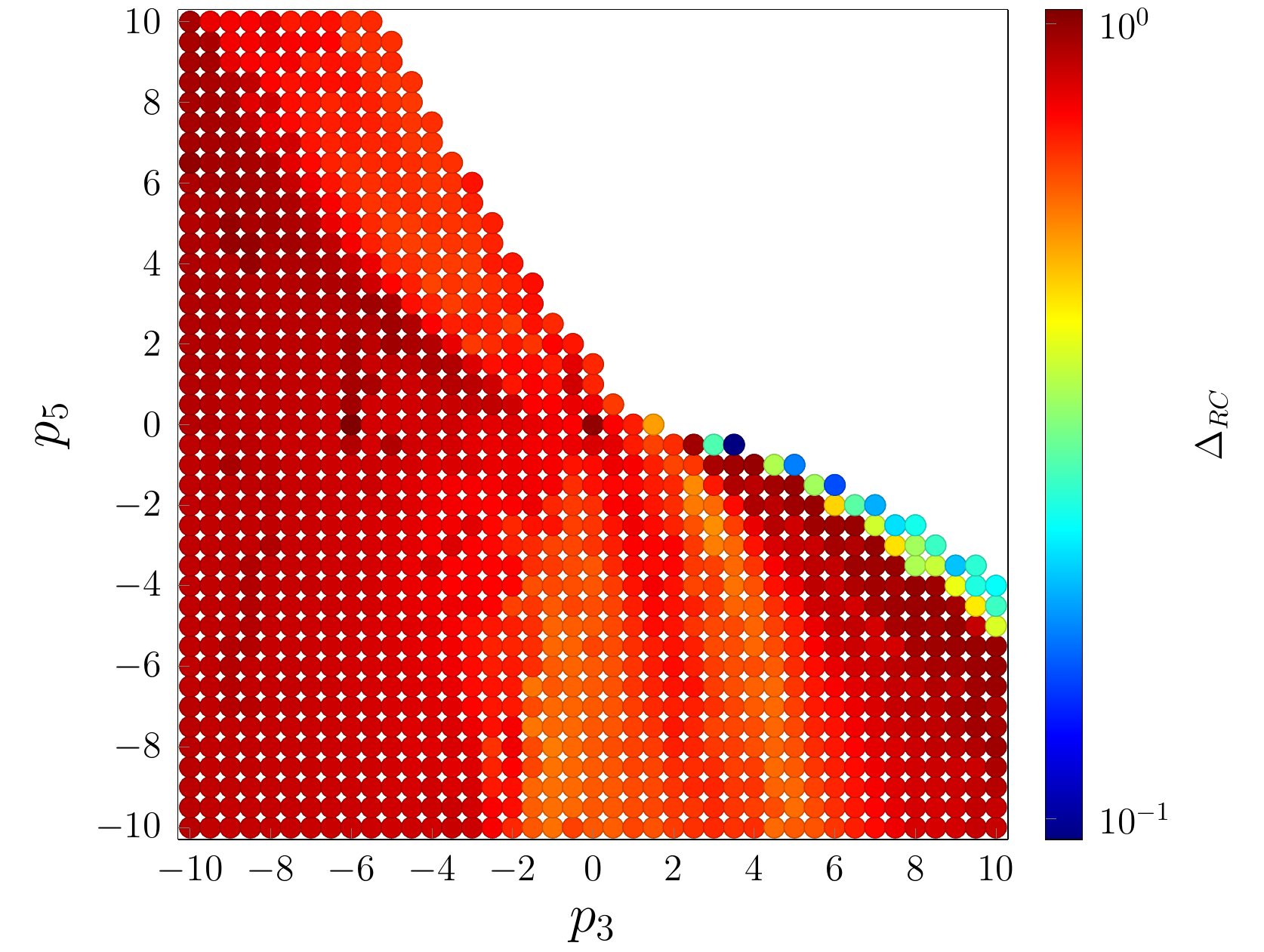} 
	\caption{\textbf{The RC performance for the case of continuous time and polynomial nodal dynamics with $f(r,\boldsymbol{\theta}) = p_1 r +p_3 r^3 + p_5 r^5$. Input and training signals are from the Lorenz attractor.}  The training error $\Delta_{RC }$ is plotted on a log scale as the parameters $p_3$ and $p_5$ are varied.  The color varies from dark blue (small error) to dark red (large error). }
	\label{fig:cont:poly5}
\end{figure}

\FS{
	We run some additional simulations to investigate the importance of carefully choosing the nonzero coefficients $p_i$ of the polynomial \eqref{general} on the reservoir performance. These are shown in Figs.\ \ref{fig:cont:poly4} and \ref{fig:cont:poly5}. The case shown in Fig.\ \ref{fig:cont:poly4} is that of a polynomial with linear, cubic and quartic order terms (but no quadratic term.) 
	Similarly to the case when linear, quadratic and cubic terms were present, we do not see the distinct boundary between sets of parameters such that the RC is globally stable (below the black line in . \ref{fig:cont:poly1}) versus sets of parameters such that $c_{\max}$ is finite (above the black curve in Fig.\ \ref{fig:cont:poly1}). %We have computed $c_{max}$ as a function of the parameters $p_3$ (coefficient of cubic term) and $p_4$ (coefficient of 4th order term.) 
	In Fig.\ \ref{fig:cont:poly4} we plot the training error  $\Delta_{RC}$ as  a  function  of  the  parameters $p_3$ and $p_4$.   The  color  represents  variations  in log scale of the training error from dark blue (small) to dark red (large). The solid black curves represent the level curves for different values of $c_{\max}$ computed from $K^*(c_{\max},\bm{\theta}) = -\alpha_{\max}(A_s)$ in the parameter space $(p_3, p_4)$. We observe that the reservoir computer performs well for those values of $p_3$  and $p_4$ for which $c_{\max} = 1$, except in the case in which $p_4=0$, which is analogous to what previously shown in Fig.\ 3 for $p_2=0$. 
	Moreover, we see for a case in which the polynomial had  first order, third order, and fifth order terms but no even power terms that the training error was always very high (shown in Fig.\ \ref{fig:cont:poly5}).  We wish to emphasize that while global stability could be achieved by a proper choice of the parameter $p_1$ alone (with all the other $p_i$, $i>1$, equal zero), our simulations show the importance of setting
	%As mentioned above, the importance of different terms in the polynomial expression appears to be an additional requirement with respect to stability of the reservoir dynamics. We conjecture that 
	certain coefficients $p_i$, $i>1$, in the polynomial \eqref{general}  to be non-zero and in particular it appears to be important to have nonzero coefficients corresponding to at least one odd power and one even power. %These  appear to be needed in order to train the system to reproduce the particular output signal one is interested in.
}

\subsection{Results for Discrete-Time Dynamics with Sigmoid  Nonlinearity}

\begin{figure}
	\centering
	\includegraphics[width=0.8\textwidth]{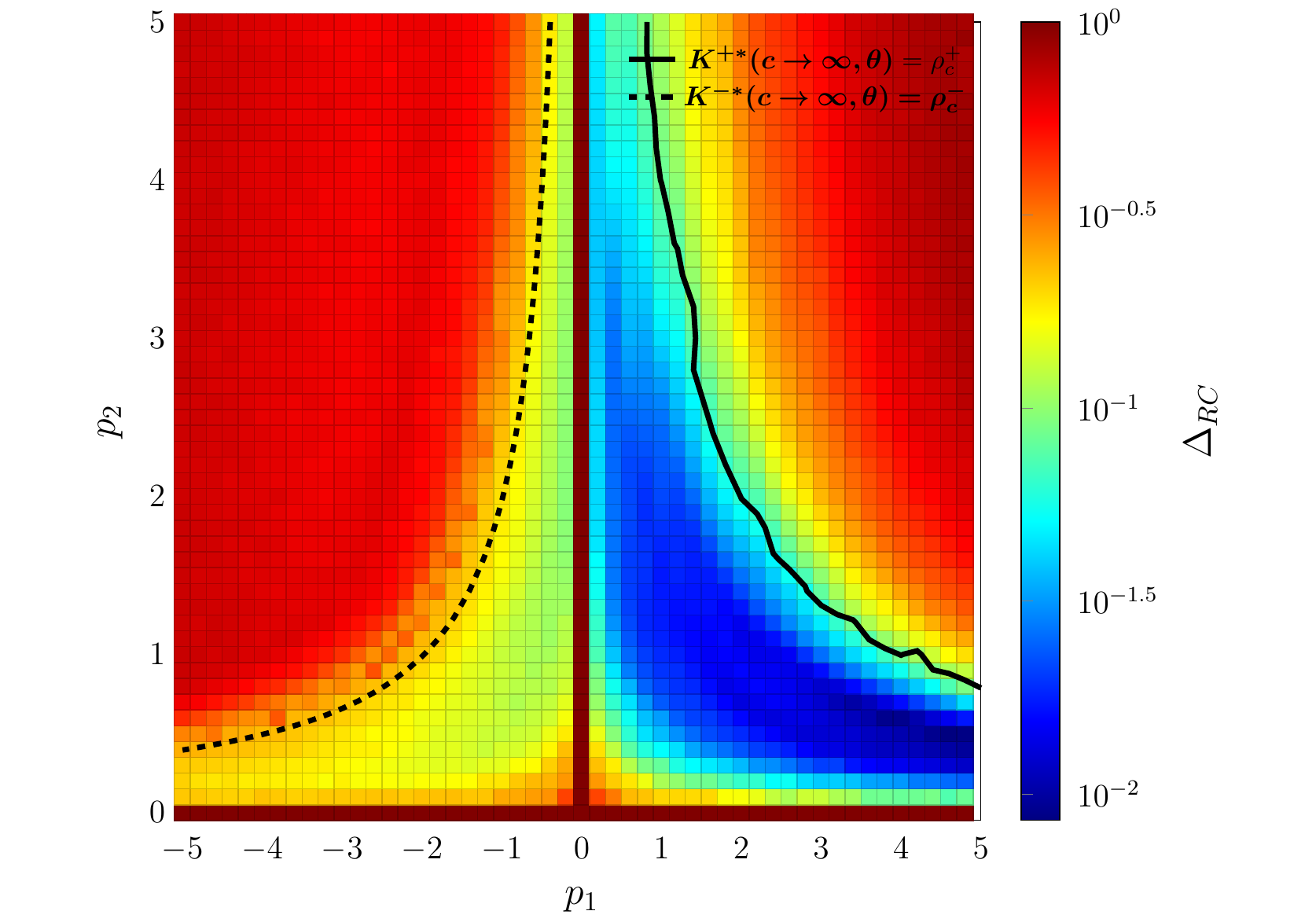}
	\caption{\textbf{Global stability and the RC performance for the case of discrete time, sigmoidal nodal dynamics. Input and training signals are from the Lorenz attractor.} The training error $\Delta_{RC }$ plotted on a log scale as a function of the parameters $p_1$ and $p_2$.  The color varies according to the training error from dark blue (small) to dark red (large). The solid black curve represents for each value of $p_1$ the value of $p_2$ such that $K^{+*}(c\rightarrow \infty,\bm{\theta}) = \rho_{c}^+$. The dashed black curve represents for each value of $p_1$  the value of $p_2$ such that  $K^{-*}(c\rightarrow -\infty,\bm{\theta}) = \rho_{c}^-$. }
	\label{fig:dis:sigmoid1}
\end{figure}

\begin{figure}
	\centering
	\includegraphics[width=0.8\textwidth]{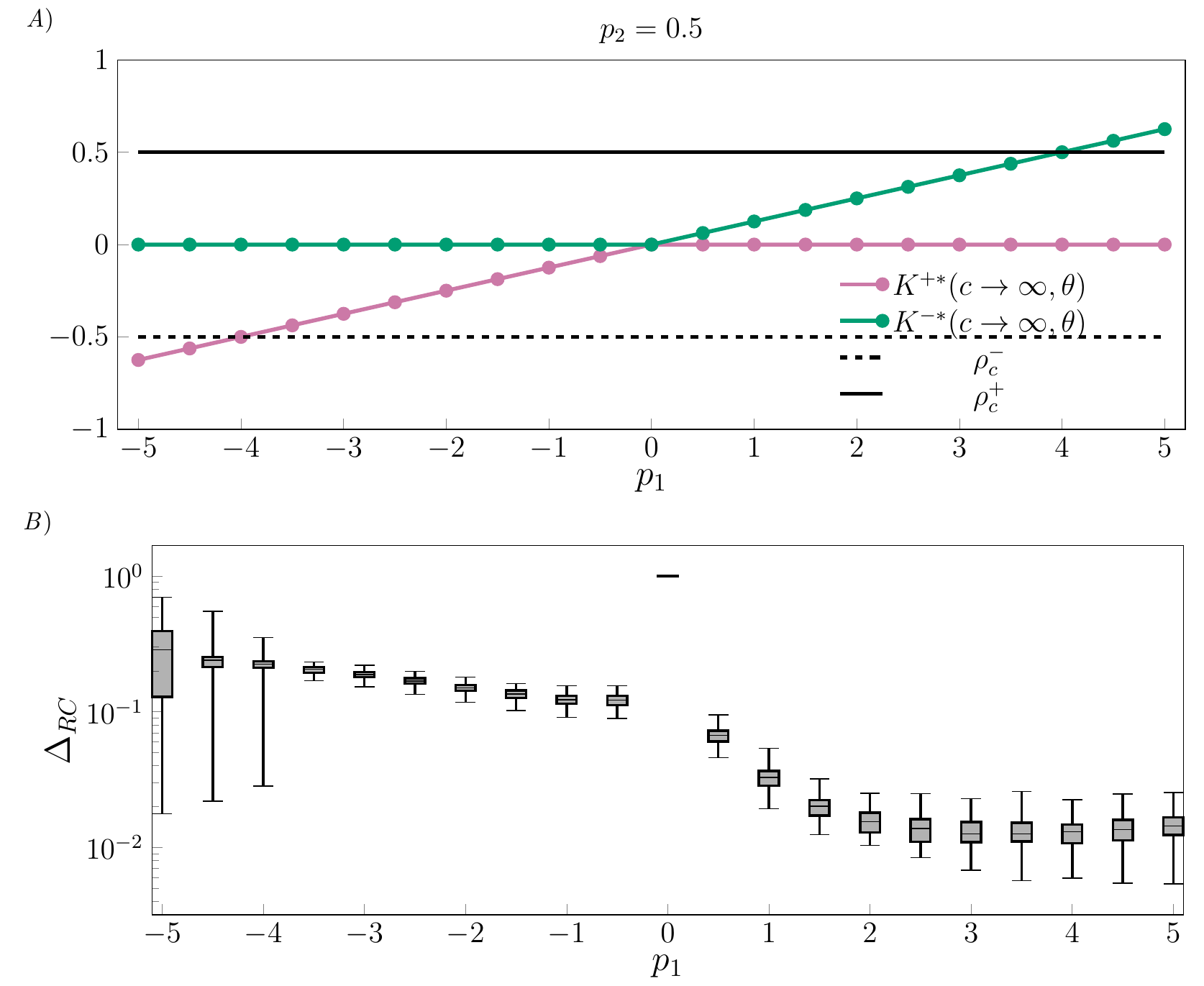}
	\caption{\textbf{Global stability and the RC performance for the case of discrete time, sigmoidal nodal dynamics (one parameter is fixed). Input and training signals are from the Lorenz attractor.}
		(\emph{A}) ${K^+}^*(c,\bm{\theta})$ (magenta)  and ${K^-}^*(c,\bm{\theta})$ (green) versus  $p_1$ for the particular value of $p_2 = 0.5$. The curves of  ${K^+}^*(c,\bm{\theta})$ and ${K^-}^*(c,\bm{\theta})$ bounded by the constant-ordinate line $\rho_{c}^+ $ (black solid) and the constant-ordinate line  $\rho_{c}^- $ (black dashed) determine the range of $p_1$ for which the system is globally stable.
		(\emph{B}) The box plot of the training error ($\Delta_{RC}$)  versus the parameter $p_1$ for $p_2 = 0.5$.}
	\label{fig:dis:sigmoid2}
\end{figure}

For the numerical  simulations of a reservoir computer with discrete-time dynamics, we set the nodal dynamics as in Eq.\  \eqref{eq:dis:dyn1} with $f(r_i,\bm{\theta}) = \frac{p_1}{ 1 + e^{-p_2 r_i}}$.
We choose the matrix $A$ as we described at the beginning of the section. We set the input signal and training signal from the Lorenz chaotic attractor and compute the training error as described in Sec.\ III.A. We consider 100 realizations of the matrix $A$ but keep the input and the training signal unchanged.
We compute the fixed point $q_i^*$ of Eq.\ \eqref{eq:dis:sigmoid} and the scalar functions $K^{-*}(c,\theta)$ and $K^{+*}(c,\theta)$ by following the Eqs.\ \eqref{eq:dis:non-hom:Km:sigmoid}-\eqref{eq:dis:non-hom:Kp:sigmoid2}.
In Fig.\ \ref{fig:dis:sigmoid1}, we plot the training error $\Delta_{RC}$ in log scale as a function of the parameters $p_1$ and $p_2$. The solid black curve represents ${K^+}^*(c\rightarrow \infty  ,\bm{\theta})= \rho^+_c$ as a function of $p_1$ and $p_2$.  The dashed black curve represents ${K^-}^*(c\rightarrow \infty  ,\bm{\theta})= \rho^-_c$ as a function of $p_1$ and $p_2$. The region between the two black curves determines the part of the parameter space $(p_1,p_2)$ for which the reservoir computer is globally stable and the reservoir computer is successful in performing the computation, while the training error may vary by different orders of magnitude as the parameter changes. On the other hand outside of this region, it is hard to assess how the system behaves. For example, the system could still be globally stable, or locally stable, in which case depending on the particular choice of input, the system dynamics might approach a different  attractor or might be driven to $\pm \infty$.
Figure  \ref{fig:dis:sigmoid2} displays the results in Fig.\ \ref{fig:dis:sigmoid1} in further detail for the cross-section $p_2 = 0.5$.
Figure\ \ref{fig:dis:sigmoid2}(\emph{A}) shows ${K^+}^*(c\rightarrow \infty  ,\bm{\theta})$ (magenta) and ${K^-}^*(c\rightarrow \infty ,\bm{\theta})$ (green) as functions of $p_1$ for constant $p_2 = 0.5$.
The black solid line is the constant-ordinate line at $\rho_c^+$ and the dashed black line is the constant-ordinate line at $\rho_c^-$. For $-4 \le p_1 \le 4$, the system is globally stable and the reservoir computer successfully performs the computation. %safe and  the performance of the reservoir computer could  vary by different orders of magnitude as the parameter changes.
In Fig.\ \ref{fig:dis:sigmoid2}(\emph{B}), we plot the training error ($\Delta_{RC}$) versus the parameter $p1$ for the particular case of $p_2 = -0.5$. 
We notice that when $-4 < p_1 < 0$, the training error is a bit high but the performance of the RC is consistence, and when  $ 0 < p_1  < 4$  the RC performs very well.

\newpage

\section{\label{sec:level4}Conclusion and Discussion}
In this paper we have used the Lyapunov design method to analyze the  nonlinear stability of a generic reservoir computer for both  continuous-time and discrete-time dynamics. \FS{Our analysis presented in the paper is simple, input independent and yet is able to predict with a certain efficacy the actual performance of the reservoir in terms of computed training error, see e.g. Fig.\ 3 and 4. Making the analysis input dependent presents the major drawback that the input may not be known a priori and one usually wants the reservoir to be able to process different input signals.}
%Under the assumptions that the input signal is normalized and scaled properly and the  weight (adjacency) matrix of the RC is properly scaled, 
Our approach allows computation of the $c({\bm \theta})$-radial region of stability about a desired fixed point, where $c$ is the radius of the stability region and ${\bm \theta}$ is a set of parameters for the nodes' individual dynamics. 
For each $c$ our approach decouples the effects of the individual nodal dynamics from those of the network topology. %A desirable property is for the RC dynamics to be  globally stable.  This is because the input signal is not known a priori and thus if the basin of attraction is some finite region, it is possible for the unknown input  to drive the system away from this basin. % of attraction.

For the case of continuous-time dynamics, we have considered a general form of polynomial nonlinearity.  We have derived a scalar function $K^*(c,{\bm \theta})$ which determines the region within the parameter space for which the system is globally stable and the reservoir performance is typically enhanced. Moreover, we have found  that the particular type of nonlinearity \textit{matters}. It is known from the literature that a RC requires nonlinearity to perform well, see e.g., \cite{jaeger2004harnessing}. Here we have found  additional evidence that \FS{ (i) the performance of a reservoir typically worsens when one of the $p_i$ coefficients in the polynomial is set to zero and (ii) it is usually important to have nonzero coefficients corresponding to at least one odd power (e.g., linear) and one even power (e.g., quadratic or quartic order.) These observations hold for both cases that the input and training signals are generated by the Lorenz and the Duffing attractors.  } 

For the case of discrete-time dynamics, a sigmoid function is used for the nodal dynamics. In this case, two scalar functions $K^{-*}(c,{\bm \theta})$ and $K^{+*}(c,{\bm \theta})$ determine the region on the parameter space for which the system is globally stable and the reservoir performance is enhanced.

Our plots in Figs.\ 3, 4, and 8 show a remarkable connection between the area of global stability in the parameter space predicted by the analysis and the performance of the RC as measured by the training error. \FS{In particular it appears that the training error worsens considerably when a change of the parameters causes loss of global stability. Our nonlinear stability analysis unveils a trade-off between the need for global stability, which is achievable by linear dynamics alone and the need for higher-order terms in the dynamics, which could in turn compromise stability.}
While fundamental insight into the exact role the nodal dynamics and the adjacency matrix have on the performance of the reservoir is an open area of research, the manual adjustment of the parameters of the reservoir computer is important to determine its dynamic regime.
Our nonlinear stability analysis  allows us to find the region within the parameter space for which  satisfactory sets of parameters may be selected.
One is able to then perform a brute search from within this region for adequate sets of parameters.
%Moreover, by a brute search inside this  parameter space, we have found the set of parameters for which improved performance can be achieved.
Moreover, by reducing the parameter space to only a finite region, this analysis empowers us to design an optimization problem to find the optimal set of parameters to maximize the performance of a reservoir computer.

	%% Include Conclusion Section
	%\section{\label{sec:level5}Conclusion}
	%\input{sections/conclusion}
	
	\section*{Acknowledgments}

	This work was supported by the National Science Foundation through Grant No. 1727948, the Office of Naval Research through Grant No. N00014-16-1-2637, and the Defense Threat Reduction Agency through Grant No. HDTRA1-12-1-0020.
	%\input{sections/end_notes}
	
	%\section*{Appendix}
	%\appendix
	%\renewcommand{\theequation}{\thesection\arabic{equation}}
	%\input{sections/appendix}

	%% Bibliography
	\bibliographystyle{unsrt}
	\bibliography{reservoir}

\end{document}